\documentclass[
pra,twocolumn,
showpacs,
superscriptaddress,
unsortedaddress,
floatfix]
{revtex4}

\usepackage{latexsym}
\usepackage{amssymb, amsmath}
\usepackage{exscale}
\usepackage{bm, graphics}
\usepackage{graphicx}

\newcommand{\abs}[1]{\lvert #1\rvert }
\newcommand{\abq}[1]{\lvert #1\rvert ^2 }
\newcommand{\D}{\mbox{\rm d}}
\newcommand{\uhb}[1]{\underline{\hat{\bf{#1}}}}
\newcommand{\hb}[1]{\hat{\bf{#1}}}
\newcommand{\uh}[1]{\underline{\hat{#1}}}

\begin{document}

\title{
QED of lossy cavities:
operator and quantum-state input-output relations
}

\author{M. Khanbekyan}
\email[E-mail address: ]{mkh@tpi.uni-jena.de}

\author{ L. Kn\"oll}

\author{D.-G. Welsch}

\affiliation{Theoretisch-Physikalisches Institut,
Friedrich-Schiller-Universit\"{a}t Jena, Max-Wien-Platz 1, D-07743
Jena, Germany}

\author{A. A. Semenov}

\affiliation{Fachbereich Physik, Universit\"{a}t Rostock,
Universit\"{a}tsplatz 3, D-18051 Rostock, Germany}
\affiliation{Institute of Physics, National Academy of Sciences of
Ukraine, 46 Prospect Nauky, UA-03028 Kiev, Ukraine}

\author{W. Vogel}

\affiliation{Institut f\"ur Physik, Universit\"{a}t Rostock,
Universit\"{a}tsplatz 3, D-18051 Rostock, Germany}

\date{\today}
\begin{abstract}
Within the framework of exact quantization of the electromagnetic
field in dispersing and absorbing media the input-output problem
of a high-$Q$ cavity is studied, with special emphasis on the
absorption losses in the coupling mirror. As expected, the cavity
modes are found to obey quantum Langevin equations, which could be
also obtained from quantum noise theories, by appropriately
coupling the cavity modes to dissipative systems, including the
effect of the mirror-assisted absorption losses. On the contrary,
the operator input-output relations obtained in this way would be
incomplete in general, as the exact calculation shows. On the
basis of the operator input-output relations the problem of
extracting the quantum state of an initially excited cavity mode
is studied and input-output relations for the $s$-parameterized
phase-space function are derived, with special emphasis on the
relation between the Wigner functions of the quantum states of the
outgoing field and the cavity field.
\end{abstract}


\maketitle

\section{Introduction}

The use of atoms interacting with light has been very
promising in handling information storage, communication, and computation
\cite{monroe:238,pellizzari:3788,knill:46,enk:205,tregenna:032305}.
In fact, optical systems do not
only allow the observation of fundamental quantum
effects \cite{hagley:1, doherty:013401, pinkse:365, hood:1447}, but
they can be also used to implement
quantum networks with photons,
which may be regarded as representing the best qubit carriers
for fast and long-distance quantum communication
\cite{pan:417, bennett:247}.
In optical systems resonatorlike devices---referred to as cavities
in the following---are indispensable elements. In particular,
high-$Q$ cavities have been well known to
offer a number of possibilities to
engineer nonclassical states of light
\cite{raimond:565, lange:063817}.

Since high-$Q$ cavities feature well-pronounced line spectra of the
electromagnetic field, which renders it possible to control
the atom-field interaction to a high degree, they are best suited
for the generation of quantum states on demand.
For example, in Ref.~\cite{law:1055} the creation of arbitrary
superposition of Fock states of light
inside a
cavity by means of controlling the time
sequence of the amplitudes and phases of the atom-field interactions of the quantized cavity
field and the external driving fields
with a trapped atom is considered.
Another proposal to generate
superposition Fock states
or coherent states in a cavity is to exploit
adiabatic interaction of
the cavity field
with an
atomic system
by achieving the transfer of ground-state Zeeman coherence onto
the cavity-mode field \cite{parkins:1578}.
It is worth noting that the idea of the
generation of a bit-stream of single photons on demand in
an optical cavity
is based on the concept of adiabatic passage
\cite{law:2067, hennrich:4872, kuhn:067901}.
In the specific context of
microwave cavities, a scheme proposed in \cite{Domokos:1}
for the
generation of photon number states on demand via $\pi$ pulse
interaction of single two-level atoms passing through a cavity,
has been experimentally realized
d\cite{brattke:3534, walther:s418}.

To measure the quantum state of a cavity field, various schemes
have been considered. As has been demonstrated experimentally
\cite{nogues:054101, bertet:200402}, the quantum state of a
microwave cavity field can be reconstructed by employing the
dispersive interaction of a single circular Rydberg atom with the
cavity field \cite{lutterbach:2547}. To reconstruct the quantum
state of a cavity field by measuring the field escaping from the
cavity, the proposal has been made to use pulsed homodyne
detection and an operational definition of the Wigner function in
terms of appropriately chosen collective mode operators
\cite{santos:033813}.

However, to further use intracavity-generated photonic quantum
states in applications
such as
quantum
networks connecting distant quantum processors and memories, the
issue of extraction
from the cavities of the quantum states
needs to be
considered very carefully with respect to quantum decoherence. In
the scheme in Refs.~\cite{cirac:3221, enk:2659} qubits that are
stored in the internal states of cold atoms, located in the
antinodes of a standing wave of a high-$Q$ optical cavity are
mapped onto photon number states, which play the role of a
communication channel by leaking out of the cavity and being
caught in a second cavity. Further, schemes for the generation of
entangled states of individual atoms held in distant cavities have
been considered (see, e.g., Refs.~\cite{browne:067901,
clark:177901, difidio:105}). In all the schemes the absorption losses
unavoidably occurring in the processes of exit and entrance of the
photons through the coupling mirrors are typically disregarded.
However, even if from the point of view of classical optics these
unwanted losses are very small, so that they effectively do not
influence classical light, they can lead to a drastic degradation
of nonclassical light features indispensable for quantum
communication (see, e.g., Ref.~\cite{scheel:063811}).

Roughly speaking, there have been two routes of treating the
input-output problem of a leaky cavity. In the first---the
quantum stochastic approach to the problem---standard Markovian
damping theory is employed, where the dynamical system is
identified with a chosen mode of the perfect cavity
(
$Q$ $\!\to$ $\!\infty$), the dissipative system is
identified with the continuum of modes outside the cavity, and a
bilinear coupling energy between the modes of the two systems is
assumed \cite{collett:3761}. In this way, the cavity mode is found
to obey a quantum Langevin equation, and operator input-output
relations
can be
derived. The
theory
can be used, e.g.,
to relate correlation functions of the outgoing field to
correlation functions of the cavity field and the incoming field
\cite{gardiner:3761} or to describe the coupling of modes of two
cavities through their
respective input and output ports
\cite{gardiner:2269,
carmichael:2273}.

In the second route---the quantum field theoretical approach to
the problem---the calculations are based on Maxwell's equations
and exact quantization of the electromagnetic field in the
presence of nonabsorbing cavity walls described in terms of
appropriately chosen real permittivities \cite{knoell:3803,
dutra:063805, viviescas:013805}. Having established the
equivalence of the two routes, thereby constructing the
interaction energy between the cavity field and the outer field,
one may try to include unwanted absorption losses in the theory by
allowing for further dissipative systems in such a way that
appropriately chosen
interaction energies
between them and the cavity modes
are added to the Hamiltonian
used in the quantum stochastic approach \cite{viviescas:013805,
khanbekyan:043807}. As we will show within the frame of exact
quantum electrodynamics in dispersing and absorbing media, this
simple concept, though leading to the correct
quantum Langevin equations for the cavity field, does not lead to
the correct operator input-output relations in general, because
the absorption losses in the coupling mirrors are not properly
taken into account.

If the operator input-output relations are known, the correlation
functions of the outgoing field can be expressed in terms of
correlation functions of the cavity field and the incoming field.
In this context the question of the calculation of outgoing-field
quantum state as a whole arises. Starting with the correct
operator input-output relations that include both wanted and
unwanted losses, we will calculate
the quantum state of
the pulselike field escaping from a high-$Q$ cavity, assuming
that the quantum state of the corresponding cavity field
at some initial time is known.

The paper is organized as follows. In Sec.~\ref{sec2} some basic
equations are given and the cavity model is introduced. The
intracavity field and the outgoing field, including the operator
input-output relations, are studied in Secs.~\ref{sec3} and
\ref{sec4}, respectively. The problem of extraction of an
initially prepared cavity-quantum state is considered in
Sec.~\ref{sec6}, and a summary and concluding remarks are given in
Sec.~\ref{sec9}. Some derivations are given in appendices.

%%%%%%%%%%%%%%%%%%%%%%%%%%%%%%%%%%%%%%%%%%%%%%%%%%%%%%%%%%%%%%%%%%%%%
%%%%%%%%%%%%%%%%%%%%%%%%%%%%%%%%%%%%%%%%%%%%%%%%%%%%%%%%%%%%%%%%%%%%%

\section{Preliminaries}
\label{sec2}

%%%%%%%%%%%%%%%%%%%%%%%%%%%%%%%%%%%%%%%%%%%%%%%%%%%%%%%%%%%%%%%%%%%%%%

\subsection{
Quantization scheme
}
\label{sec2.1}

Let us consider $N$ atoms (with the $A$th atom being at
position $\mathbf{r}_A$) that
in electric dipole approximation interact with the electromagnetic
field in the presence of linear dielectric media of spatially
varying and frequency-dependent complex permittivity
\begin{equation}
    \label{1.29}
      \varepsilon({\bf r},\omega) = \varepsilon'({\bf r},\omega)
      + i\varepsilon''({\bf r},\omega).
\end{equation}
Note that due to causality
the real and imaginary parts \mbox{$\varepsilon'({\bf r},\omega)$}
and \mbox{$\varepsilon''({\bf r},\omega)$}, respectively, are
uniquely related to each other through the Kramers-Kronig
relations.
Following
the approach to quantization of the macroscopic
Maxwell field as given in
Refs.~\cite{gruner:1818,scheel:700,knoell:1,scheel:4094,ho:053804},
we may write the multipolar-coupling
Hamiltonian
in the form of
\cite{text}
\begin{equation}
   \label{1.1}
        \hat{H} =
        \hat{H}_\mathrm{field} + \hat{H} _\mathrm{atom}
        + \hat{H} _\mathrm{int}.
\end{equation}
Here,
\begin{equation}
   \label{1.3}
        \hat{H}_\mathrm{field} =
        \int\! \D^3{r} \int_0^\infty\! \D\omega
      \,\hbar\omega\,\hb {f}^{\dagger}({\bf r },\omega)\cdot
      \hb{ f}({\bf r},\omega)
\end{equation}
is the Hamiltonian of the system composed of the electromagnetic
field and the medium, including a reservoir necessarily
associated with material absorption, with
\mbox{$\hb{ f}({\bf r},\omega)$}
[and \mbox{$\hb{f}^\dagger({\bf r},\omega)$}]
being bosonic fields that play the role of the dynamical
variables of the composed system,
\begin{equation}
    \label{1.11}
      \bigl[\hat{f}_{\mu} ({\bf r}, \omega),
      \hat{f}_{\mu'} ^{\dagger } ({\bf r }',  \omega ') \bigr]
      = \delta _{\mu \mu'}\delta (\omega - \omega  ')
      \delta ^{(3)}({\bf r} - {\bf r }') ,
\end{equation}
\begin{equation}
    \label{1.13}
      \bigl[\hat {f}_{\mu} ({\bf r}, \omega),
      \hat {f}_{\mu'}  ({\bf r }', \omega ') \bigr] = 0
\end{equation}
(the Greek letters label the Cartesian components).
Further,
\begin{equation}
 \label{1.5}
      \hat{H}_\mathrm{atom} =
      \sum _A \sum _k
        \hbar \omega _{Ak} \hat{S} _{Akk}
\end{equation}
is the atomic Hamiltonian and
\begin{equation}
 \label{1.7}
         \hat{H} _\mathrm{int}=
        -\sum _A
        \hb{ d}_A\cdot
        \hb{E}({\bf r}_A)
\end{equation}
is the (multipolar-)interaction energy, where
\begin{equation}
   \label{1.6}
   \hat{S} _{Ak'k} =
   | k'\rangle_A {_A}\!\langle k |
\end{equation}
are the flip operators of the $A$th atom,
\begin{equation}
   \label{1.8}
    \hb{ d}_A = \sum _{kk'}
    {\bf d} _{Akk'}  \hat{S} _{Akk'}
\end{equation}
is its electric dipole moment
($ {\bf d} _{Akk'}$ $\!=$ $\!{_A}\!\langle k|
 \hb{ d}_A | k' \rangle_A$),
and $\hb{E}({\bf r})$ is the medium-assisted electric field,
which expressed in terms of $\mathbf{f}(\mathbf{r},\omega)$ [and
$\mathbf{f}^\dagger(\mathbf{r},\omega)$] reads
\begin{equation}
    \label{1.19}
      \hb{E}({\bf r}) = \int_0^\infty \D\omega\,
      \uhb{E}({\bf r},\omega) + \mbox{H.c.}\,,
\end{equation}
\begin{align}
      \label{1.37}
&      \uhb{ E}({\bf r},\omega)
      = i \mu _0 \sqrt{\frac {\hbar \epsilon _0}{\pi}}\,
      \omega^2
      \nonumber\\&\hspace{2ex}\times
      \int \D^3r'\sqrt{\varepsilon''({\bf r}',\omega)}\,
      \mathsf{G}({\bf r},{\bf r}',\omega)
      \cdot\hb{f}({\bf r}',\omega)
\end{align}
where
the classical Green tensor $\mathsf{G}({\bf r},{\bf r}',\omega)$,
which also corresponds to the quantum field-theoretical
retarded Green tensor (see, e.g., Ref.~\cite{Abrikosov}),
is the solution to the equation
\begin{equation}
      \label{1.39}
      \bm{\nabla}\!\times\!
    \bm{\nabla}\!  \times \mathsf{G}  ({\bf r }, {\bf r }', \omega)
      - \frac {\omega ^2 } {c^2} \,\varepsilon ( {\bf r } ,\omega)
      \mathsf{G}  ({\bf r}, {\bf r }', \omega)
      =  \bm{\delta} ^{(3)}  ({\bf r }-{\bf r }')
      \end{equation}
together with the boundary condition
\begin{equation}
    \label{1.29-1}
    \mathsf{G}  ({\bf r}, {\bf r }', \omega)
    \to 0 \ \mathrm{for}\ |\mathbf{r}-\mathbf{r} '| \to \infty
\end{equation}

It is not difficult to see, that
in the Heisenberg picture $\hb{ f}({\bf r },\omega,t)$
obeys the
equation of motion
\begin{align}
   \label{1.30}
&     \dot{\hat{ {\bf f} }}
     ({\bf r},\omega, t)
     = \frac {1} {i\hbar} \bigl[ \hb{ f}({\bf r},\omega, t)
       , \hat {H}\bigr]
     = -i \omega \hb{ f}({\bf r},\omega, t)
\nonumber\\&\hspace{2ex}
     + \mu_0 \omega ^2 \sqrt{\frac{\varepsilon_0}{\hbar \pi}}\,
     \sqrt{\varepsilon''({\bf r},\omega)}\,
     \sum_A  \hb{ d}_A (t)\cdot \mathsf{G} ^*({\bf r}_A,{\bf r},\omega)
     ,
\end{align}
which after formal integration leads to
\begin{equation}
 \label{1.32}
\hat{ {\bf f} } ({\bf r},\omega, t)
= \hat{{\bf f}}_\mathrm{free}({\bf r},\omega, t)
+ \hat{{\bf f}}_\mathrm{s}({\bf r},\omega, t),
\end{equation}
where
\begin{equation}
      \label{1.15}
        \hat{{\bf f}}_\mathrm{free}({\bf r},\omega, t)
        = e^{-i \omega (t-t')}\hat{{\bf f}}_\mathrm{free}({\bf r},\omega, t')
\end{equation}
and
\begin{align}
  \label{1.32-1}
&    \hat{ {\bf f} }_\mathrm{s} ({\bf r},\omega, t)
    =   \mu_0 \omega ^2 \sqrt{\frac{\varepsilon_0}{\hbar \pi}}\,
     \sqrt{\varepsilon''({\bf r},\omega)}
\nonumber\\&\hspace{1.5ex}\times
   \sum_A \!
   \int\! \D t'\, \Theta(t\!-\!t')
   \hb{ d}_A (t')\!\cdot\!\mathsf{G} ^*({\bf r}_A,{\bf r},\omega)
   e^{-i\omega (t-t')}.
\end{align}
Substitution of Eq.~(\ref{1.32}) together with Eqs.~(\ref{1.15}) and
(\ref{1.32-1}) into Eq.~(\ref{1.37}) yields the corresponding
source-quantity representation of
$\underline{\hat{\mathbf{E}}}(\mathbf{r},\omega, t)$.

%%%%%%%%%%%%%%%%%%%%%%%%%%%%%%%%%%%%%%%%%%%%%%%%%%%%%%%%%%%%%%%%%%%%%%%%%
\subsection{Cavity model}
\label{sec2.2}

\begin{figure}[t]
\includegraphics{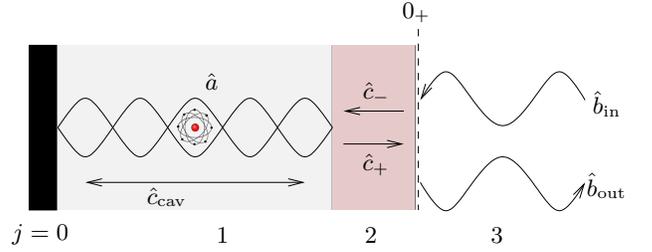}
\caption{\label{fig}Scheme of the cavity. The fractionally
transparent mirror [region (2)] is modeled by a dielectric plate.
The active sources are in region 1, which can also contain some
medium.}
\end{figure}
For the sake of transparency, let us consider a one-dimensional
cavity modeled by a planar dielectric 4-layer system
(Fig.~\ref{fig}). In particular, the layers \mbox{$j$ $\!=$ $\!0$}
and \mbox{$j$ $\!=$ $\!2$}, respectively, are assumed to
correspond to perfectly and fractionally reflecting mirrors which
confine the cavity (layer \mbox{$j$ $\!=$ $\!1$}). In what follows
we use, with respect to $z$, shifted coordinate systems such that
\mbox{$0$ $\!<$ $\!z$ $\!<$ $\!l$} for \mbox{$j$ $\!=$ $\!1$},
\mbox{$0$ $\!<$ $\!z$ $\!<$ $\!d$} for \mbox{$j$ $\!=$ $\!2$}, and
\mbox{$0$ $\!<$ $\!z$ $\!<\infty$} for \mbox{$j$ $\!=$ $\!3$}.
Applying the one-dimensional version of Eq.~(\ref{1.37}) together
with Eqs.~(\ref{1.32})--(\ref{1.32-1}) to the field in the $j$th
layer of permittivity $\varepsilon_j(\omega)$ \mbox{($j$ $\!=$
$\!1,2,3$)}, we may write
\begin{equation}
\label{2.0}
\uh{ E}{^{(j)}}(z, \omega,t)
= \uh{ E}^{(j)}_\mathrm{free}(z, \omega,t)
+ \uh{ E}^{(j)}_\mathrm{s}(z, \omega,t),
\end{equation}
with
\begin{align}
\label{1.45-1}
&\uh{ E}^{(j)}_\mathrm{free}(z, \omega,t)
= i\omega\mu_0
      \nonumber\\&\hspace{2ex}\times
      \sum _{j'=1}^3 \int _{[j']} \D z'\,
      G ^{(jj')}(z, z', \omega)
      \uh{j}^{(j')}_\mathrm{free}(z',\omega,t)
\end{align}
and
\begin{align}
  \label{1.40}
&    \underline{\hat{ E}}^{(j)}_\mathrm{s} (z, \omega, t) =
    \frac{i}{\pi \epsilon _0
      \mathcal{A}}\,
    \frac{\omega ^2} {c^2}
    \sum_A \int \D t' \,\Theta(t-t')
\nonumber \\&\hspace{8ex}\times\,
   e^{-i\omega (t-t')}
   \hat{ d}_A (t')\, \mathrm{Im}\,G ^{(1j)}(z_A, z,\omega)
\end{align}
($\mathcal{A}$, mirror area), where the integral
relation (\ref{a4.3}) has been employed.
Here,
$[j']$ indicates integration over the $j'$th layer,
the abbreviating notation
\begin{equation}
      \label{1.47}
      \uh { j}^{(j)}_\mathrm{free} (z, \omega,t ) =
      \omega\, \sqrt { \frac {\hbar \epsilon _0} {\pi {\cal A} }\,
      \varepsilon '' _j (\omega) } \,
      \hat { f}^{(j)}_\mathrm{free} (z, \omega,t )
\end{equation}
is used, and it is assumed that the active atomic sources
are localized inside the cavity. The (nonlocal part of
the) Green function reads \cite{khanbekyan:063812}
\begin{align}
      \label{1.49}
&      G^{(jj')}(z, z', \omega )
\nonumber\\&\hspace{2ex}
      = {\textstyle\frac{1}{2}} i \bigl[
      \mathcal{E} ^{(j)>}    (z, \omega )\,
      \Xi^{jj'}    \mathcal{E} ^{(j')<}    (z', \omega)
      \Theta (j-j')
\nonumber \\ &\hspace{4ex}
      + \mathcal{E}^{(j)<}    (z, \omega )\,
      \Xi^{j'j}   \mathcal{E} ^{(j')>}   (z', \omega )
      \Theta (j'-j)
      \bigr] ,
\end{align}
where
the functions
\begin{equation}
      \label{1.51}
\mathcal{E} ^{(j)>}    (z, \omega ) =
       e^{i \beta _j (z-d _j)}
      + r  _{j/3}
      e^{-i \beta _j (z-d _j)}
      \end{equation}
and
\begin{equation}
      \label{1.53}
      {\mathcal{E}} ^{(j)<}    (z,   \omega )\, =\,
        e^{-i \beta _j z}
      + r  _{j/0}   e^{i \beta _j z},
\end{equation}
respectively, represent waves of unit strength traveling
rightward and leftward in the $j$th layer and being reflected at
the boundary [note that $\Theta (j$ $\!-$ $\!j')$  means
\mbox{$\Theta (z$ $\!-$ $\!z')$} for $j$ $\!=$ $\!j'$]. Further,
$\Xi^{jj'}$ is defined by
\begin{equation}
      \label{1.55}
      \Xi^{jj'} =
      \frac{1}{\beta_3 t _{0/3}}\,
      \frac{t _{0/j}e^{ i \beta _j d _j}}{D _{ j}}\,
      \frac{t _{3/j'}e^{ i \beta _{j'} d _{j'}}}{D _{ j'}}\,,
      \end{equation}
where
\begin{equation}
      \label{1.57}
      D_{j} = 1 - r _{j/0} r _{j/3} e^{2 i \beta _j d _j}
      \end{equation}
and
\begin{align}
      \label{1.59}
      \beta _j &\equiv
      \beta _j(\omega) =
      \sqrt {\varepsilon _j (\omega)} \,\frac {\omega} {c}
      \nonumber\\
      &= [n_j ' (\omega) + i n_j '' (\omega) ]\,\frac {\omega} {c}
      = \beta _j ' + i \beta _j ''
      \quad  ( \beta_j ', \beta_j '' \geq 0 )
      \end{align}
($d_1$ $\!=$ $\!l$, $d_2$ $\!=$ $\!d$,
$d_3$ $\!=$ $\!0$).
The quantities \mbox{$t_{j/j'}$ $\!=$
$\!(\beta_j/\beta_{j'})t_{j'/j}$} and $r_{j/j'}$ denote,
respectively, the transmission and reflection coefficients between
the layers $j'$ and $j$, which can be recursively determined (for
recursion formulas, see Appendix \ref{app.1}).

%%%%%%%%%%%%%%%%%%%%%%%%%%%%%%%%%%%%%%%%%%%%%%%%%%%%%%%%%%%%%%%%%%%%%%%%%
%%%%%%%%%%%%%%%%%%%%%%%%%%%%%%%%%%%%%%%%%%%%%%%%%%%%%%%%%%%%%%%%%%%%%%%%%

\section{Cavity Field}
\label{sec3}

To further evaluate the equations given above, we first consider
the field inside the cavity ($j$ $\!=$ $\!1$). In order to make
contact with the familiar standing-wave expansion in the idealized
case of a lossless cavity, it is useful to rewrite the
equations with the aim to obtain a nonmonochromatic mode expansion
that takes into account the finite line widths due to the
wanted input-output coupling
and
the unwanted
absorption losses that unavoidably exist in practice.

%%%%%%%%%%%%%%%%%%%%%%%%%%%%%%%%%%%%%%%%%%%%%%%%%%%%%%%%%%%%%%%%%%%%%

\subsection{Nonmonochromatic mode expansion}
\label{sec3.1}

We begin with the free field.
{F}rom Eqs.~(\ref{1.45-1}) and (\ref{1.49}) it follows that
$\underline{\hat{ E}}^{(1)}_\mathrm{free} (z, \omega,t)$ can
be represented in the form of
\begin{align}
\label{2.1}
&    \underline{\hat{E}}_\mathrm{free}^{(1)} (z, \omega,t) =
\nonumber \\&\hspace{1ex}
        \frac {1}{D_1}
        \left[e^{i \beta_1 z} \!+\! r_{13} e^{-i \beta_1 (z-2 l)}\right]
        \left[
\hat{C}
        ^{(1)}_{<+} (z, \omega,t)
        \!-\!
\hat{C}
        ^{(1)}_{<-} (z, \omega,t)\right]
\nonumber \\&\hspace{3ex}
        -\,\frac{2 i\sin( \beta_1 z)}{D_1}
        \Bigl\{
\hat{C}
        ^{(1)}_{>-} (z, \omega,t)
        + r_{13} e^{2 i \beta _1 l}
\hat{C}
        ^{(1)}_{>+} (z, \omega,t)
\nonumber \\&\hspace{10ex}
        +\,\frac {t_{21}e^{i \beta _1 l}}
        {D_2'}\,
        \left[
\hat{C}
        ^{(2)}_{-} ( \omega,t)
        + r_{23} e^{2i \beta _2 d}
\hat{C}
        ^{(2)}_{+} ( \omega,t)\right]
\nonumber \\&\hspace{10ex}
        +\,t_{31}
        e^{i \beta _1 l}
\hat{C}
        ^{(3)}_{-} ( \omega,t)
        \Bigr\},
\end{align}
where
\begin{align}
\label{2.5}
&
\hat{C}
    ^{(1)}_{<\pm} (z, \omega,t)
\nonumber\\&\hspace{2ex}
    = -\frac{ \mu _0 c}{2 n_1}
    \int_{[1]} \D z'\,\Theta(z\!-\!z')
    e^{ \mp i \beta _1 z'}
     \uh { j}^{(1)}_\mathrm{free} (z ' ,\omega ,t),
\\[.5ex]
\label{2.7}
&
\hat{C}
    ^{(1)}_{>\pm} (z, \omega,t)
\nonumber\\&\hspace{2ex}
    =-\frac{ \mu _0 c}{2 n_1}
    \int_{[1]} \D z'\,\Theta(z'\!-\!z)
    e^{ \mp i \beta _1 z'}
     \uh { j}^{(1)}_\mathrm{free} (z ' ,\omega, t),
\end{align}
\begin{align}
\label{2.9}
&
\hat{C}
    ^{(2)} _{\pm} (\omega, t )
    =-\frac{ \mu _0 c}{2 n_2}
    \int_{[2]} \D z'\,
    e^{\mp i \beta _2 z'}
     \uh { j}^{(2)}_\mathrm{free} (z', \omega, t) ,
\\[.5ex]
\label{2.11}
&
\hat{C}
    ^{(3)}_{-} (\omega,t)
    =-\frac{ \mu _0 c}{2 n_3}
    \int_{[3]} \D z'\,
    e^{i \beta _3 z'}
    \underline{\hat { j}}^{(3)}_\mathrm{free} (z', \omega, t) ,
\end{align}
and
\begin{equation}
      \label{2.3}
D_2'(\omega) = 1- r_{21} r_{23} e^{2 i \beta_2 d}.
\end{equation}
Inspection of Eq.~(\ref{2.1}) shows that the function
$D_1(\omega)$ defined by Eq.~(\ref{1.57}) for $j=1$
characterizes the spectral response of the cavity. In particular,
its zeros determine the (complex) resonance frequencies $\Omega_k$,
\begin{equation}
      \label{2.15}
 D_1 (\Omega _{k})
    = 1+ r_{13}(\Omega _{k})
    e^{2 i \beta_1 (\Omega _{k}) l} = 0.
\end{equation}
Note that when the coupling mirror is not a single plate
but---as in practice---a multilayer system, then $r_{13}(\omega)$
is the reflection coefficient of the multilayer system
and Eq.~(\ref{2.15}) applies as well.
Decomposing $\Omega_{k}$ into real and
imaginary parts according to
\begin{equation}
      \label{2.17}
      \Omega _{k} = \omega_{k}
      - {\textstyle\frac {1} {2}}i\Gamma _{k},
\end{equation}
we can write the formal solution to
Eq.~(\ref{2.15}) in the form of
\begin{align}
  \label{2.19.3}
  \omega _{k}  &=  \frac{c}{l}\frac{1}{2|n_1|^2}
\nonumber\\&\quad\times
  \biggl\{ \!n_1'
  \biggl[2 \pi k
  \!+ \!\pi \!- \!\tan ^{-1}\!\biggl(\frac{r_{13}''}{r_{13}'}\biggr)
  \!\biggr] \!- \!n_1 '' \ln|r_{13}|\!\biggr\}
\end{align}
and
\begin{align}
  \label{2.19.5}
  \Gamma _{k}  &= \frac{c}{l}\frac{1}{|n_1|^2}
\nonumber\\&\quad\times
  \biggr\{\!n_1''\biggl[2\pi k \!+ \!\pi \!-
  \tan ^{-1}\!\biggl(\frac{r_{13}''}{r_{13}'}\biggr) \!\biggr]
  \!+ \!n_1 ' \ln|r_{13}|\!\biggr\}
\end{align}
[$n_1$ $\!=$ $\!n_1(\Omega_{k})$,
$r_{13}$ $\!=$ $\!r_{13}(\Omega_{k})$], from which
$\omega _{k}$ and $\Gamma _{k}$ may be calculated by
iteration, by starting, e.g., with the resonance frequencies
of the lossless cavity.

Let $s(t)$ be a function of time whose Fourier transform
is given by
\begin{equation}
\label{2.21}
\underline{s}(\omega) =
\frac{S(\omega)}{D_1(\omega)}= \int \mathrm{d}t\,e^{i\omega t}s(t)
\end{equation}
and assume that $S(\omega)$ is analytic in the lower half-plane.
Employing the residue theorem, we may write
\begin{equation}
  \label{2.23}
s(t) = \!\int\!\frac{\mathrm{d}\omega}{2\pi}\,e^{-i\omega t}
\frac{S(\omega)}{D_1(\omega)} =
  \sum _k \frac{c}{2 n_1 l}\,\Theta (t)
  e^{-i\Omega _{k} t} S ( \Omega _{k}).
\end{equation}
Applying Eqs.~(\ref{2.21}) and
(\ref{2.23}) to the $c$-number functions
$\sin( \beta_1 z)D^{-1}$,
$\sin( \beta_1 z)r_{13} e^{2 i \beta _1 l}D^{-1}$, and
$[e^{i \beta_1 z}$ $\!+$ $\!r_{13} e^{-i \beta_1 (z-2 l)}]D^{-1}$
in Eq.~(\ref{2.1})
and disregarding (irrelevant high-frequency)
contributions that may arise from poles other than those
of $D_1^{-1}(\omega)$, we may rewrite
$\underline{\hat{ E}}_\mathrm{free}^{(1)}(z, \omega, t)$ as
\begin{equation}
\label{2.25-0}
\underline{\hat{ E}}_\mathrm{free}^{(1)}
(z, \omega, t)
= \sum _k \underline{\hat{ E}}_{k\mathrm{free}}^{(1)}
(z, \omega, t),
\end{equation}
with
\begin{align}
      \label{2.25}
& \underline{\hat{ E}}_{k\mathrm{free}}^{(1)}
(z, \omega, t) =
i
\sqrt{\frac{\mu_0c\hbar\omega }{\pi\mathcal{A} n_1}}
       \frac{c}{2 n_1\!(\Omega _k) l}\,\sin [\beta_1\!(\Omega _k) z]
\nonumber \\&\hspace{2ex}\times\,
       \int \D t'\,e^{-i\Omega_k(t-t')}\Theta (t-t')
\nonumber \\&\hspace{7ex}
\times\, \Bigl[
      T (\omega )\hat {b}_{\mathrm{in}}(\omega, t')
    + \sum _{\lambda }
    A_{\lambda}(\omega )\hat {c}_{\lambda}\!(\omega, t')\Bigr]
\end{align}
($\lambda$ $\! =$  $\!\mathrm{cav },+,-$),
where
the operators
$\hat{c}_\lambda(\omega,t)$ are defined according to
\begin{align}
\label{2.27}
&\hat{c}_\mathrm{cav}(\omega, t) =
    -\alpha_\mathrm{cav} \sqrt{\frac{\pi\mathcal{A}}{\mu_0c\hbar\omega}}
    \frac{ \mu _0 c}{2 n_1}
\nonumber\\&\hspace{12ex}\times\,
    \int_{[1]} \D z\,
    \sin(\beta _1 z)
     \uh{ j}^{(1)}_\mathrm{free} (z,\omega, t),
\\[.5ex]
\label{2.29}
&    \hat{c}_{\pm} (\omega, t) =
    \alpha _{\pm}\sqrt{\frac{\pi\mathcal{A}}{\mu_0c\hbar\omega}}
\nonumber\\&\hspace{12ex}\times\,
    \left[
        e^{i \beta _2 d}
        \hat{C}
        ^{(2)} _+ (\omega, t)
        \pm
\hat{C}
^{(2)}_- (\omega, t)
        \right],
\\[.5ex]
\label{2.31}
&        \hat{b}_{\mathrm{in}}(\omega, t) =
    \frac{2|n_3|}{\sqrt{n_3'}}
    \sqrt{\frac{\pi\mathcal{A}}{\mu_0c\hbar\omega}}\,
\hat{C}
^{(3)} _{-} (\omega, t),
\end{align}
with
\begin{align}
  \label{2.33}
   \alpha_\mathrm{cav} &= \alpha_\mathrm{cav} (\omega)
\nonumber\\
&= 2\sqrt{2}|n_1|
    \left[n_1 '\sinh(2 \beta _1'' l)
    - n_1'' \sin(2 \beta _1 'l)\right]^{-\frac{1}{2}} ,
\\
\label{2.35}
    \alpha_{\pm} &= \alpha _{\pm}(\omega)
\nonumber\\
&= |n_2|e^{ \beta _2'' d /2}
    \left[n_2'\sinh(\beta _2'' d)\pm n_2''\sin(\beta _2'd)
    \right]^{-\frac{1}{2}},
\end{align}
\begin{align}
\label{2.57}
&
A _{ \mathrm{cav}}(\omega ) = -4 i \,\frac{\sqrt{n_1
}}{\alpha _\mathrm{cav}}
,
\\[.5ex]&
  \label{2.59}
A _{\pm}(\omega ) =
  - \frac {t_{21}\sqrt{n_1
}} {D_2'\alpha_{\pm }}
  \left(r_{23} e^{i \beta _2 d} \pm 1\right)
  e^{
i \beta _1 l} ,
\\[.5ex]&
  \label{2.61}
T (\omega ) = - \frac{t_{31}\sqrt{n_1
n_3'} }{|n_3|}
  \,e^{
i \beta _1 l}.
\end{align}
It is straightforward to prove that the operators
$\hat{c}_\lambda(\omega,t)$
satisfy Bose commutation
relations:
\begin{align}
\label{5.11}
&\bigl[\hat{c}_\lambda(\omega, t),
\hat{c}_{\lambda'}^\dagger(\omega', t')\bigr]
= \delta_{\lambda\lambda'}\delta (\omega - \omega ')
          e^{-i \omega (t-t')},
\\
\label{5.11-1}
&\bigl[\hat{b}_\mathrm{in}(\omega, t),
\hat{b}_\mathrm{in}^\dagger(\omega', t')\bigr]
= \delta (\omega - \omega ')
          e^{-i \omega (t-t')},
\end{align}
with all other commutators being zero.

To calculate the electric free field
\begin{equation}
      \label{2.41-0}
      \hat{E}_\mathrm{free}^{(1)}(z, t) = \int_0^\infty
      \mathrm{d}\omega\,\underline{\hat{E}}_\mathrm{free}^{(1)}(z,\omega,t)
      + \mathrm{H.c.}
\end{equation}
[cf. Eq.~(\ref{1.19})],
we subdivide the $\omega$ axis
into intervals
\mbox{$(\Delta_k)$ $\!\equiv$
$[\frac{1}{2}(\omega_{k-1}$ $\!+$
$\!\omega_k),\frac{1}{2}(\omega_k$ $\!+$ $\!\omega_{k+1})]$}
and write
\begin{equation}
      \label{2.41}
      \hat{E}_\mathrm{free}^{(1)}(z, t)
      = \sum _k \hat{E}_{k\mathrm{free}}^{(1)}(z,t) + \mathrm{H.c.},
\end{equation}
where
\begin{equation}
      \label{2.43}
      \hat{E}_{k\mathrm{free}}^{(1)}(z, t)
      = \int_{(\Delta_k)} \D\omega\,
      \underline{\hat{E}}_\mathrm{free}^{(1)}(z,\omega,t)
\end{equation}
(recall that the index $k$ is used to numerate the resonances
of the cavity).
Substitution of Eq.~(\ref{2.25-0}) together with Eq.~(\ref{2.25})
into Eq.~(\ref{2.43}) yields
\begin{align}
\label{2.43-1}
\hat{E}_{k\mathrm{free}}^{(1)}(z, t) = &\int_{(\Delta_k)} \D\omega\,
      \underline{\hat{E}}_{k\mathrm{free}}^{(1)}(z,\omega,t)
\nonumber\\&
+ \sum_{k'\neq k} \int_{(\Delta_k)} \D\omega\,
\underline{\hat{E}}_{k'\mathrm{free}}^{(1)}(z,\omega,t) .
\end{align}
For sufficiently high-$Q$ cavities,
i.e., $\Gamma_{k}$ $\!\ll $
$\!\Delta\omega_k$,
with
\mbox{$\Delta\omega_k$
$\!=$ $\!\frac{1}{2}
(\omega_{k+1}$ $\!-$ $\!\omega_{k-1})$}
being the width of the $k$th interval,
the second term in Eq.~(\ref{2.43-1}) can be regarded as being
small compared with the first one and may be omitted in general,
leading to
\begin{equation}
\label{2.43-2}
\hat{E}^{(1)}_{k\mathrm{free}}(z, t)
=
\int_{(\Delta_k)} \D\omega\,
      \underline{\hat{E}}_{k\mathrm{free}}^{(1)}(z,\omega,t).
\end{equation}
In this approximation, Eq.~(\ref{2.41}) reduces to
\begin{equation}
\label{2.43-3}
\hat{E}_\mathrm{free}^{(1)}(z, t) = \sum _k \int_{(\Delta_k)} \D\omega\,
      \underline{\hat{E}}_{k\mathrm{free}}^{(1)}(z,\omega,t).
    \end{equation}
Note that within the approximation scheme used, the
lower (upper) limit of integration in Eq.~(\ref{2.43-3}) may be
extended to $-\infty$ ($+\infty$).

To determine the  source field
\begin{equation}
      \label{2.41-1}
      \hat{E}_\mathrm{s}^{(1)}(z, t) = \int_0^\infty
      \mathrm{d}\omega\,\underline{\hat{E}}_\mathrm{s}^{(1)}(z,\omega,t)
      + \mathrm{H.c.}
,
\end{equation}
we start with Eq.~(\ref{1.40}) together with Eq.~(\ref{1.49}).
Performing the Fourier transformation and using the resonance
properties of the cavity response function
[see Eqs.~(\ref{2.15})--(\ref{2.23})], we obtain, in the same approximation
that leads from Eq.~(\ref{2.41-0}) to Eq.~(\ref{2.43-3}),
\begin{equation}
\label{2.50-0}
\hat{E}_\mathrm{s}^{(1)}(z, t) = \sum _k
      \hat{E}_{k\mathrm{s}}^{(1)}(z,t)
      + \mathrm{H.c.}
      ,
\end{equation}
where
\begin{align}
  \label{2.50}
&    \hat{E}^{(1)}_{k\mathrm{s}}(z, t) =
    \frac{i \omega _k\sin [\beta _1(\Omega _k) z]}
     {\epsilon _0
       \varepsilon _1 \!(\Omega _k) l
       \mathcal{A}
     }
    \sum_A \int\! \D t'\, \Theta(t\!-\!t')
\nonumber \\&\hspace{6ex}
    \times\,
   e^{-i\Omega_k (t-t')}
   \hat{ d}_A (t')
  \sin [\beta _1\!(\Omega _k) z_A]  + \mathrm{H.c.}
  \,.
\end{align}
Note that in Eq.~(\ref{2.50}) it is assumed that
$e^{i\omega_kt}\hat{d}_A(t)$ may be regarded as being
an effectively slowly varying quantity.
Combination of $\hat{E}_\mathrm{free}^{(1)}(z, t)$ and
$\hat{E}_\mathrm{s}^{(1)}(z, t)$ to the full intracavity field
\begin{equation}
\label{2.50-1}
\hat{E}^{(1)}(z, t) = \hat{E}_\mathrm{free}^{(1)}(z, t)
+ \hat{E}_\mathrm{s}^{(1)}(z, t)
\end{equation}
yields the nonmonochromatic mode expansion sought.

%%%%%%%%%%%%%%%%%%%%%%%%%%%%%%%%%%%%%%%%%%%%%%%%%%%%%%%%%%%%%%%%%%%%%%%%%
\subsection{Quantum Langevin equations}
\label{sec3.2}

To bring Eq.~(\ref{2.50-1})
[with $\hat{E}_\mathrm{free}^{(1)}(z, t)$ from Eq.~(\ref{2.43-3})
together with Eq.~(\ref{2.25}) and
$\hat{E}_\mathrm{s}^{(1)}(z, t)$ from Eq.~(\ref{2.50-0})
together with Eq.~(\ref{2.50})]
in a more familiar form, we introduce
the operators
\begin{align}
\label{2.45}
&        \hat{c}_{k \lambda}( t)
        = \frac{ 1}{\sqrt{2 \pi}}
        \int_{(\Delta_k)} \D\omega\,
        \hat{c}_{\lambda}(\omega, t),
\\
\label{2.49}
&        \hat{b}_{k\mathrm{in}}( t)
        = \frac{ 1}{\sqrt{2 \pi}}
        \int_{(\Delta_k)} \D\omega\,
        \hat{b}_{\mathrm{in}}(\omega, t),
\end{align}
which, on a time scale $\Delta t$ $\!\gg$
$\Delta\omega_k^{-1},\Delta\omega_{k'}^{-1}$,
obviously obey [recall Eqs.~(\ref{5.11}) and (\ref{5.11-1})]
the commutation relations
\begin{align}
\label{5.13}
&\bigl[\hat{c}_{k \lambda} (t),
\hat{c}_{k' \lambda'}^\dagger(t')\bigr]
= \delta _{kk'} \delta_{\lambda\lambda'}\delta (t - t'),
\\
\label{5.13-1}
&\bigl[\hat{b}_{k \mathrm{in}}(t),
\hat{b}_{k' \mathrm{in}} ^\dagger(t')\bigr]
= \delta _{kk'} \delta (t - t').
\end{align}
Further, recalling Eqs.~(\ref{2.25}), (\ref{2.43-3}),
(\ref{2.50-0}), and (\ref{2.50}), we may rewrite
Eq.~(\ref{2.50-1}) as
($\Gamma_{k}$ $\!\ll $
$\!\Delta\omega_k$)
\begin{equation}
      \label{2.51}
      \hat{ E}^{(1)}(z, t) = \sum_k
      E_k(z)
      \hat{a}_k (t) + \mathrm{H.c.},
\end{equation}
where the standing wave mode functions are defined as
\begin{equation}
  \label{2.53}
  E_k (z) = i\omega _k
  \left[\frac {\hbar}
  { \epsilon_0 \varepsilon _1( \omega_k) l {\cal A} \omega_k}
  \right]^{\frac{1}{2}}\sin[\beta_1(\omega_k) z],
\end{equation}
and
\begin{align}
\label{2.55}
&\hat{a}_k(t) =
        \int \D  t' \, \Theta ( t - t')
        e^{-i\Omega_{k}(t-t')}
\nonumber \\&\hspace{5ex}\times
\biggl\{
\biggl[ \frac {c} {2 n_1(\omega_k) l}\biggr]^{\frac{1}{2}}
\biggl[
T_k \hat{b}_{k\mathrm{in}} ( t') +
  \sum _{\lambda }
    A _{k {\lambda }} \hat{c}_{k {\lambda }} ( t')
\biggr]
\nonumber \\&\hspace{10ex}
      -\frac{i}{ \hbar }
      \sum_A
      E_k (z_A)
      \hat{ d}_A (t')
  \biggr\}
\end{align}
[$T_k$ $\!=$ $\!T(\omega_k)$,
$A_{k \lambda}$ $\!=$ $\!A_{\lambda}(\omega_k)$].

{F}rom Eq.~(\ref{2.55}) it is not difficult to see that $\hat{a}_k$
obeys the Langevin equation
\begin{align}
    \label{2.71}
&        \dot{\hat{a}}_k (t)
        = - i\left(\omega_{k}
        - {\textstyle\frac{1}{2}}i\Gamma_k\right) \hat{a}_k (t)
        - \frac{i}{ \hbar}
        \sum_A
        E_k (z_A)
        \hat{ d}_A(t)
\nonumber\\&\hspace{5ex}
+ \biggl[ \frac {c} {2 n_1(\omega_k) l}\biggr]^{\frac{1}{2}}
\biggl[ T_k \hat{b}_{k\mathrm{in}} ( t)
+ \sum _{\lambda } A _{k\lambda } \hat{c}_{k \lambda } ( t)
\biggr] ,
\end{align}
and it can be proved (see Appendix \ref{app.2.3}) that
the equal-time commutation relation
\begin{equation}
  \label{5.9}
  [\hat{a}_k(t),\hat{a}_{k'}^{\dagger}(t)] = \delta_{kk'}
\end{equation}
holds.

The damping rate in the first term
on the right-hand side of Eq.~(\ref{2.71}) can be decomposed
as follows (see Appendix \ref{app.2.1}):
\begin{align}
\label{5.1}
&        \Gamma_k  =
         \gamma_{k\mathrm{rad}}
        +
        \gamma_{k\mathrm{abs}}\,,
\\
\label{5.3}
&     \gamma_{k\mathrm{rad}}
     = \frac{c}{2 |n_1(\omega_k)| l}   |T_k|^2,
\\
\label{5.7}
&     \gamma_{k\mathrm{abs}} =
    \sum _{\lambda }\gamma_{k\lambda} =
     \frac{c}{2 |n_1(\omega_k)| l}
    \sum _{\lambda }|A_{k\lambda }|^2 .
\end{align}
Here, $\gamma_{k\mathrm{rad}}$ is the radiative decay rate
describing the transmission losses due to the input-output
coupling and $\gamma_{k\mathrm{abs}}$ is the (nonradiative) decay
rate describing the absorption losses inside the cavity (term
proportional to $|A_{k\mathrm{cav}}|^2$) and inside the mirror (terms
proportional to $|A_{k\pm}|^2$). Accordingly,
the Langevin noise force
as given by the third term on the right-hand side of Eq.~(\ref{2.71})
consists of the contributions associated with the losses due
to the input-output coupling [term proportional to $T_k\hat {b}
_{k\mathrm{in}}(t)$] and the absorption losses inside the cavity
[term proportional to $A_{k\mathrm{cav}}\hat {c}_{k\mathrm{cav}}(t)$]
and inside the mirror [terms proportional to $A_{k\pm}\hat
{c}_{k\pm}(t)$].

Equation (\ref{2.71}) can be regarded as a generalization
of the results derived in Ref.~\cite{knoell:3803}
for a leaky cavity without material absorption to a realistic cavity
which gives rise to both radiative
and unwanted (nonradiative) absorption losses.
In particular, when $\varepsilon_1(\omega_k)$ can be regarded as
being real,
then the second term on the right-hand side of Eq.~(\ref{2.71})
is nothing but the familiar commutator term $(i\hbar)^{-1}[\hat{a}_k,
\hat{H}_\mathrm{int}]$, where
\begin{equation}
\label{2.73}
\hat{H}_\mathrm{int}
= - \sum_A \sum_k  E_k(z_A)\hat{d}_A \hat{a}_k + \mathrm{H.c.}.
\end{equation}
Moreover, from Eq.~(\ref{2.71}) together with
Eqs.~(\ref{5.1})--(\ref{5.7}) it is seen that the effect of
absorption losses on the intracavity field may be equivalently
described within the frame of Markovian damping theory, with
\begin{align}
\label{2.74}
&\hat{H}_{\mathrm{abs,int}}
= \hbar \sum_\lambda \sum_k \!\int_{(\Delta_k)} \!\!\D\omega\,
   \biggl[ \frac {c} {2 n_1(\omega) l}\biggr]^{\frac{1}{2}}
   A_{k\lambda}(\omega)\hat{a}_k^\dagger\hat{c}_{k\lambda}(\omega)
\nonumber\\&\hspace{10ex}
   + \mathrm{H.c.}
\end{align}
being the total interaction energy between the cavity modes and
the dissipative systems responsible for absorption. Thus
Eq.~(\ref{2.71}) can be also regarded as an extension of the
results derived in Ref.~\cite{gardiner:3761} within the frame of
quantum noise theories, by adding to the Hamiltonian therein an
interaction energy of the type (\ref{2.74}). Needless to say that
also other than the dissipative channels considered here can be
included in the interaction energy. The unwanted losses attributed
to the cavity wall that has been assumed to be perfectly
reflecting is a typical example. Note that Eq.~(\ref{2.74})
implies that each cavity mode is coupled to its own dissipative
systems.

%%%%%%%%%%%%%%%%%%%%%%%%%%%%%%%%%%%%%%%%%%%%%%%%%%%%%%%%%%%%%%%%%%%%%%%%%%%
\section{Field outside the
cavity}
\label{sec4}

Once the cavity field is expressed in terms of nonmonochromatic
mode operators $\hat{a}_k(t)$, the question arises of how the
outgoing field is related to it. To answer, we first rewrite the
outgoing field using Eqs.~(\ref{2.0})--(\ref{1.40}) with $j$ $\!=$
$\!3$ and proceed similarly as in the case of the cavity field.

\subsection{
Outgoing field
}
\label{out.1}
%%%%%%%%%%%%%%%%%%%%%%%%%%%%%%%%%%%%%%%%%%%%%%%%%%%%%%%%%%%%%%%%%%%%%%%%%%

We
again begin with the free
field. Inserting the Green tensor as given by Eq.~(\ref{1.49}) in
Eq.~(\ref{1.45-1}) ($j$ $\!=$ $\!3$) and separating the incoming
and outgoing parts propagating along $-z$ and $z$, respectively,
we may represent the outgoing part at $z$ $\!=$ $\!0^+$ (cf.
Fig.~\ref{fig}) as follows (see Appendix \ref{app.1.3}):
\begin{align}
      \label{3.3}
&        \underline{\hat{ E}}_{{\rm out}, {\rm free}} ^{(3)}
      (z, \omega, t)
        \bigr|_{z=0^+}
\nonumber \\&
        = \frac {t_{13}e^{i \beta_1 l}}{ D_1 }
    \biggl\lbrace
    \left[
        \hat {C} ^{(1)}_{<+}(l, \omega, t )
        -
        \hat {C} ^{(1)}_{<-}(l, \omega, t )
    \right]
\nonumber \\&\hspace{5ex}
        - \frac {t_{21} e^{i \beta_1 l} } { D_2 '}
    \left[
        r_{23} e^{2i \beta_2 d}
        \hat {C} ^{(2)}_{+}(\omega, t )
        +
        \hat {C} ^{(2)}_{-}(\omega, t )
    \right]
\nonumber \\&\hspace{28ex}
    -\,t_{31} e^{i \beta_1 l}
    \hat {C} ^{(3)}_{-}(\omega, t )
      \biggr\rbrace
 \nonumber \\&
 +\,\frac {t_{23} e^{i \beta_2 d}} {  D_2 '}
    \left[
        \hat {C} ^{(2)}_{+}(\omega, t )
        +
        r_{21}
        \hat {C} ^{(2)}_{-}(\omega, t )
    \right]
        \!+\! r_{31}
    \hat {C} ^{(3)}_{-}(\omega, t )
,
 \end{align}
where $\hat {C} ^{(1)}_{<\pm}(l, \omega, t )$ is defined
by Eq.~(\ref{2.5}) (for $x$ $\!=$ $\!l$), and
$\hat {C} ^{(2)}_{\pm}(\omega, t )$
and $\hat {C} ^{(3)}_{-}(\omega, t )$ are defined by Eqs.~(\ref{2.9})
and (\ref{2.11}), respectively.
Treating the term $ D_1^{-1}(\omega )$ in the same
way as that leading from Eq.~(\ref{2.1}) to Eq.~(\ref{2.25-0})
together with Eqs.~(\ref{2.25})--(\ref{2.35}), from Eq.~(\ref{3.3})
we derive
\begin{equation}
\label{3.5-0}
\underline{\hat{ E}}_{{\rm out}, {\rm free}}^{(3)}
(z, \omega, t)
\bigr|_{z=0^+}
= \sum _k  \underline{\hat{ E}}
    _{k  \mathrm{
      out}, {\rm free}}^{(3)}
(z, \omega, t)
\bigr|_{z=0^+},
\end{equation}
where
\begin{align}
      \label{3.5}
&    \underline{\hat{E}} _{k\mathrm{out}, {\rm free}}^{(3)}
    (z, \omega, t)\bigr|_{z=0^+}
    =
    \frac{1}{2}
\sqrt{\frac{\mu_0c\hbar\omega }{\pi\mathcal{A} n_1}}
    \frac{c}{2 n_1\!(\Omega _k) l}
               \,
               t_{13}
\nonumber \\[.5ex]&\hspace{4ex}\times\,
     \int \D t'
     e^{-i\Omega _ {k}(t-t')}
       \Theta (t-t')
      \, e^{i \beta _1 l}
\nonumber \\[.5ex]&\hspace{6ex}\times\,
     \biggl[ T (\omega )\hat {b}_{\mathrm{in}}(\omega, t')
    + \sum _{\lambda }
    A_{\lambda}(\omega )\hat {c}_{\lambda}\!(\omega, t') \biggr]
\nonumber \\[.5ex]&\hspace{2ex}
    +\frac {1}{2}\,
    \sqrt{\frac{\mu_0c\hbar\omega}{\pi\mathcal{A}}}
    \,\biggl\lbrace\frac{\sqrt{n_3'} r_{31}}{|n_3|}\,
    \hat {b} _{k\mathrm {in}} (\omega, t)
\nonumber \\[2ex]&\hspace{6ex}
    + \frac {t_{23}  } {  D_2 '}
    \biggl[\frac{r_{21}e^{ i \beta_2 d}+1}{\alpha _+ }
    \,\hat {c}_{k+}(\omega, t)
\nonumber \\[.5ex]&\hspace{15ex}
    - \frac{r_{21}e^{ i \beta_2 d}-1}{\alpha _- }
    \,\hat {c}_{k-}(\omega, t)\biggr]\biggr\rbrace .
\end{align}
As we see from this equation,
there are three physically different contributions
to the outgoing free field. The first (integral)
term proportional to $t_{13}$ represents the fraction
of the cavity field transmitted through the mirror
[cf. Eq.~(\ref{2.25})]. The term
proportional to $r_{31}\hat {b} _{k\mathrm{in}}(\omega, t)$
represents the reflected part of the incoming field,
whereas the terms proportional to $t_{23} \hat{c} _{k\pm}(\omega, t)$
describe the field attributed to
the noise sources inside the mirror.

Integrating Eq.~(\ref{3.5}) with respect to $\omega$, we
obtain the
free-field part of the outgoing electric field
in the time-domain,
\begin{equation}
\label{3.5.1}
\hat{E} _{{\rm out}, {\rm free}}^{(3)} (z, t)\bigr|_{z=0^+} =
\sum _k \hat{E}_{k {\rm
      out}, {\rm free}}^{(3)}(z,t)\bigr|_{z=0^+}
+ \mathrm{H.c.}
\end{equation}
[cf. Eqs.~(\ref{2.41-0}) and (\ref{2.41})],
where, within the approximation scheme used,
\begin{equation}
\label{3.5.3}
\hat{E}_{k {\rm out}, {\rm free}}^{(3)}
(z, t)\bigr|_{z=0^+}
=
\int_{(\Delta_k)} \D\omega\,
     \underline{\hat{E}}_{k {\rm out}, {\rm free}}^{(3)}
(z,\omega,t)\bigr|_{z=0^+}
\end{equation}
[cf. Eqs.~(\ref{2.41})--(\ref{2.43-3})].

Starting from Eq.~(\ref{1.40}) ($j$ $\!=$ $\!3$)
together with Eq.~(\ref{1.49}), we may rewrite the source-field part
of the outgoing electric field to obtain,
in close analogy to Eqs.~(\ref{2.41-1}) and (\ref{2.50}),
\begin{align}
\label{3.51}
\hat{ E}_{{\rm out}, {\rm s}}^{(3)}(z,t)
\bigr|_{z=0^+}
&= \hat{ E}_{{\rm s}}^{(3)}(z,t)
\bigl|_{z=0^+}
\nonumber\\
&= \sum _k  \hat{ E}
    _{k, {\rm s}}^{(3)}(z,t)
    \bigr|_{z=0^+} + \mathrm{H.c.},
\end{align}
where
\begin{align}
  \label{3.53}
&    \hat{E}^{(3)}_{k\mathrm{s}}(z, t)\bigr|_{z=0^+} =
    \int_{(\Delta_k)}\!\D\omega\,
    \underline{\hat{E}}^{(3)}_{k\mathrm{s}}(z,\omega,t)\bigr|_{z=0^+}
    + \mathrm{H.c.}
\nonumber\\&\hspace{2ex}
= \frac{ \omega _k t_{13}\exp [i \beta _1(\Omega _k) l]}
     {2\epsilon _0
       \varepsilon _1(\Omega _k) l
       \mathcal{A}
     }
    \sum_A \int \D t'\, \Theta(t-t')
\nonumber \\&\hspace{6ex}
    \times\,
   e^{-i\Omega_k (t-t')}
   \hat{ d}_A (t')
  \sin [\beta _1(\Omega _k) z_A]  + \mathrm{H.c.}.
\end{align}
Finally, combination of the free-field part and the source-field part
yields the full outgoing field at $z$ $\!=$ $\!0^+$,
\begin{equation}
\label{2.53-1}
\hat{E}_\mathrm{out}^{(3)}(z, t)\bigr|_{z=0^+}
= \hat{E}_\mathrm{out,free}^{(3)}(z, t)\bigr|_{z=0^+}
+ \hat{E}_\mathrm{s}^{(3)}(z, t)\bigr|_{z=0^+}.
\end{equation}

%%%%%%%%%%%%%%%%%%%%%%%%%%%%%%%%%%%%%%%%%%%%%%%%%%%%%%%%%%%%%%%%%%%%%%%%%%

\subsection{Global input-output relation}
\label{sec4.2}

Let us restrict our attention, for simplicity, to a cavity in free space, i.e.,
$n_3\to 1$,
and define
the operators
\begin{equation}
      \label{3.7}
      \hat{ b}_{ \mathrm{out}}  (\omega, t)
      =
      2
      \sqrt{\frac{\pi\mathcal{A}}{\mu_0c\hbar\omega}}
     \,\uh{ E}^{(3)} _{\mathrm{out}}(z, \omega, t)
     \bigr|_{z=0^+}
.
\end{equation}
By using the formulas given in Section \ref{out.1} it is not
difficult to see that we may rewrite the $\omega$-integrated operator
\begin{equation}\label{3.9-0}
\hat{ b}_{\mathrm{out}}(t)
= \frac{1}{\sqrt{2\pi}} \int
\mathrm{d}\omega\,
\hat{ b}_{ \mathrm{out}}(\omega,t)
\end{equation}
as
\begin{equation}
\label{3.9}
\hat{ b}_{\mathrm{out}}(t)
= \sum_k \hat{b}_{k\mathrm{out}}(t),
\end{equation}
where
\begin{equation}
\label{3.9.2}
\hat{ b}_{k\mathrm{out}}(t)
= \frac{1}{\sqrt{2\pi}} \int_{(\Delta_k)} \mathrm{d}\omega\,
\hat{ b}_{k\mathrm{out}}(\omega,t),
\end{equation}
with $\hat{b}_{k\mathrm {out}}(\omega, t)$ being given by
\begin{align}
  \label{3.20.1}
&        \hat{b}_{k\mathrm {out}}(\omega, t)
        =
        2
        \sqrt{\frac{\pi\mathcal{A}}{\mu_0c\hbar\omega}}
        \,
        \uh{ E}^{(3)}_{k s}(z, \omega, t)
        \bigr|_{z=0^+}
\nonumber \\&\hspace{1ex}
      + \frac{c}{2n_1 l}\,T_k ^{({\rm o})}(\omega)
        \int \D  t' \, \Theta ( t - t')
        e^{-i\Omega_{k}(t-t')}
\nonumber \\&\hspace{6ex}\times\,
     \biggl[
      T_k(\omega )
      \hat {b}_{k\mathrm{in}}(\omega, t')
      + \sum _{\lambda }
      A_{k\lambda}(\omega )
      \hat {c}_{k\lambda}\!(\omega, t')
      \biggr]
\nonumber\\&\hspace{1ex}
        + A _{k+} ^{({\rm o})}(\omega)\hat{c}_{k+}(\omega, t)
        + A _{k-} ^{({\rm o})}(\omega)\hat{c}_{k-}(\omega, t)
\nonumber\\&\hspace{1ex}
        + R_k ^{({\rm o})}(\omega)\hat{b}_{k\mathrm{in}} (\omega, t ).
\end{align}
Here, the functions $A_{k\pm}^\mathrm{(o)}(\omega)$,
$R_k^\mathrm{(o)}(\omega)$, and $T_k^\mathrm{(o)}(\omega)$ are
defined as follows:
\begin{align}
  \label{3.17}
&  A_{k\pm} ^{({\rm o})}(\omega) =
  \frac {t_{23}} {
  D_2'}
  \frac{1\pm r_{21} e^{i \beta _2 d}}{\alpha_{\pm}}\,,
\\
  \label{3.13}
&  T_k^{({\rm o})}(\omega) =
  \frac{ t_{13}} {
   \sqrt{n_1}}
\, e^{i \beta _1 l}
,
\\
  \label{3.15}
&  R_k^{({\rm o})}(\omega ) =
  r_{31} .
\end{align}

Performing in Eq.~(\ref{3.9.2}) the $\omega$ integral,
on extending again the lower (upper) limit to $-\infty$
($+\infty$) and recalling Eqs.~(\ref{2.55}) and (\ref{3.53}),
we see that the source term in Eq.~(\ref{3.20.1}) and
the second (integral) term in this equation sum up to a
term proportional to the cavity-field operator $\hat{a}_k(t)$.
Thus, from Eqs.~(\ref{3.9})--(\ref{3.20.1}) it follows that
(\mbox{$\Gamma_{k}$ $\!\ll $
$\!\Delta\omega_k$})
\begin{align}
\label{3.11}
    \hat{b} _{k\mathrm {out}}(t)
    = &\left[\frac{c}{2n_1(\omega_k) l}\right]^{\frac{1}{2}}
    T_k^{({\rm o})}\hat {a} _k(t)
    + R_k ^{({\rm o})}\hat{b} _{k\mathrm{in}}(t)
\nonumber\\[2ex]&
    + A _{k+} ^{({\rm o})}\hat{c}_{k+}(t)
    + A _{k-} ^{({\rm o})}\hat{c}_{k-}(t)
\end{align}
[$T_k^{({\rm o})}$ $\!=$ $\!T_k^{({\rm o})}(\omega_k)$,
$A^{({\rm o})}_{k\pm}$ $\!=$ $\!A^{({\rm o})}_{k\pm}(\omega_k)$,
$R_k^{({\rm o})}$ $\!=$ $\!R_k^{({\rm o})}(\omega_k)$].

If on the right-hand side in Eq.~(\ref{3.11}) the
third term and the forth term, which result from
the absorption losses in the coupling mirror,
are omitted, then Eq.~(\ref{3.11}) reduces to
the well-known input-output relation \cite{gardiner:3761,knoell:3803}
for a leaky cavity whose losses
solely result from the wanted radiative input-output
coupling, in which case $|R_k^{({\rm o})}(\omega_k)|$ $\!=$ $\!1$ holds.
It can be shown that the operators
that appear in Eq.~(\ref{3.11})
obey commutation relations of the type given in
Ref.~\cite{knoell:3803}. In particular
on the time scale considered,
i.e., $\Delta t$ $\!\gg$
$\Delta\omega_k^{-1},\Delta\omega_{k'}^{-1}$,
the commutation relation
\begin{equation}
\label{5.21} \bigl[
        \hat {b} _{k{\rm out}}  ( t) ,
        \hat {b}^{\dagger} _{k'{\rm out}} ( t')
         \bigr]
         = \delta _{kk'} \delta (t-t ')
\end{equation}
holds (see Appendix \ref{app.2.2}).

As we know from Sec.~\ref{sec3.2}, the damping of the cavity modes
due to unwanted losses can simply be described by introducing into
the Hamiltonian an interaction energy of the type (\ref{2.74}) and
treating its effect in Markovian approximation. However, since
such a simple interaction energy does not allow to include in the
theory effects such as
for example the influence of
the coupling-mirror-assisted absorption
on the outgoing field via the \emph{incoming} field,
the last two terms in Eq.~(\ref{3.11})
are
missing and hence
\mbox{$|R_k^\mathrm{(o)}|$ $\!=$ $\!1$} is set
(see, e.g., Ref.~\cite{gardiner:3761}).
As a matter of fact, due to unavoidable losses it is always
observed that
$|R_k^\mathrm{(o)}|$ $\!<$ $\!1$ in practice.
The additional noise associated with these losses
is just described by the last two terms in
Eq.~(\ref{3.11}).

The
input-output relation (\ref{3.11})
can be used, e.g., to calculate
correlation functions of the outgoing field in terms of (in general
mixed) correlation functions of the cavity field, the incoming field,
and the dissipative channels. In what follows we will not consider
the one or the other correlation function, but focus on the quantum
state as a whole. In this connection it
should be stress laid on the fact that
Eq.~(\ref{3.11}) as a global input-output relation
does not specify the incoming and outgoing (nonmonochromatic)
modes as well as those of the dissipative channels which are really
connected in the relevant frequency interval defined by the
bandwidth of the cavity mode and hence
basically carry the quantum state.

%%%%%%%%%%%%%%%%%%%%%%%%%%%%%%%%%%%%%%%%%%%%%%%%%%%%%%%%%%%%%%%
\section{Quantum state of the
outgoing field}
\label{sec6}

For the sake of transparency let
us suppose that
during the passage of atoms through the cavity
the $k$th cavity mode
is prepared
in some quantum state and assume that the preparation time
is sufficiently short compared with the decay time $\Gamma_k^{-1}$,
so that the two time scales are well distinguishable.
In this case we may assume that at some time $t_0$
(when the atom leaves the cavity)
the cavity mode is prepared in a given quantum state and
its evolution in the further course of time (i.e., for times
\mbox{$t$ $\!\ge$ $\!t_0$})
can be treated
as free-field evolution.
To specify the relevant modes,
it is useful not to use Eq.~(\ref{3.11}) but
return to Eq.~(\ref{3.20.1})
and relate therein $\hat{b}_{k\mathrm {out}}(\omega, t)$
to $\hat{a}_k(t_0)$.
It can be proved (see Appendix \ref{app.7}) that
on the (relevant) time scale
\mbox{$\Delta t$ $\!\gg$ $\!\Delta\omega_k^{-1}$}
Eq.~(\ref{3.20.1}) can be rewritten as
\begin{align}
  \label{3.21}
&        \hat{b}_{k \mathrm {out}}
        (\omega, t)
\nonumber\\[.5ex]&\hspace{1ex}
         = \,
        \left[\frac{c}{2n_1(\omega_k) l}\right]^{\frac{1}{2}}
    T_k ^{({\rm o})}
    \,\frac{1} {\sqrt{2\pi}}
        \int _{t_0}^{t+\Delta t }
    \!
    \D t ' \,
        e^{-i\omega (t  - t')}
    \hat{a} _k (t')
\nonumber\\[.5ex]&\hspace{2ex}
     + A _{k+} ^{({\rm o})}\hat{c}_{k+}(\omega, t_0)
        e^{-i\omega (t - t_0)}
        +
       A _{k-} ^{({\rm o})}\hat{c}_{k-}(\omega, t_0)
        e^{-i\omega (t - t_0)}
\nonumber\\[.5ex]&\hspace{2ex}
        + R_k^{({\rm o})}\hat{b}_{k \mathrm{in}} (\omega, t _0)
        e^{-i\omega (t - t_0)} .
\end{align}
Note that integration of both sides of
Eq.~(\ref{3.21}) with respect to $\omega$ over the interval
$(\Delta_k)$
leads to the input-output relation (\ref{3.11})
($n_3\to 1$).
It should be mentioned that, for the special case of purely
radiative losses, an equation of the type of Eq.~(\ref{3.21}) could
be also found from the quantum stochastic theory in
Ref.~\cite{gardiner:3761}, which would however suggest that
its validity only requires the condition
\mbox{$\Delta t$ $\!>$ $\!0$} to be satisfied.

Substituting Eq.~(\ref{2.55}) (for $t$ $\!\ge$ $t_0$)
together with Eqs.~(\ref{2.45})
and (\ref{2.49})
into Eq.~(\ref{3.21}), we derive
\begin{equation}
\label{3.23}
        \hat{b}_{k \mathrm {out}}
        (\omega, t) =
        F^*_k (\omega, t)
        \hat{a}_k(t_0)
        + \hat{B}_k(\omega,t),
\end{equation}
where the $c$-number function $F_k(\omega,t)$ reads
\begin{align}
\label{3.25}
        F_k (\omega, t) &=
         \frac{i} {\sqrt{2\pi}}
        \left(\frac{c}{2 n_1^*  l}\right)^{1/2} \!
        T_k^{({\rm o})*} \,
        e^{i\omega (t-t_0)}
\nonumber\\[.5ex]&\quad
\times \,
\frac
         {\exp \left[-i (\omega - \Omega ^* _k) (t+\Delta t-t_0)
                \right] -1
         }
         {\omega - \Omega ^* _k}
\, ,
\end{align}
and the operator $\hat{B}_k(\omega,t)$ is
a linear functional of the operators
$\hat{b}_{k\rm{in}}(\omega,t_0)$
and $\hat{c} _{k\lambda}(\omega,t_0)$:
\begin{align}
  \label{3.27}
&  \hat{B}_k(\omega,t) =
      \int _{(\Delta_k)}
       \!\D\omega '\,
\Bigl[
G_{k\rm{in}}^*
(\omega, \omega ', t)
       \,\hat{b}_{k\mathrm {in}} (\omega ', t_0)
\Bigr.
\nonumber\\[2ex]&\hspace{5ex}
\Bigl.
+ \sum_{\lambda}
G_{k\lambda}^*
(\omega, \omega ', t)
 \,\hat{c} _{k\lambda} (\omega',t_0)
\Bigr].
\end{align}
Here,
\begin{align}
  \label{3.29}
&
G_{k\rm{in}}
        (\omega, \omega ', t) =
        T_k ^{({\rm o})*} T_k^*
        \upsilon _k (\omega, \omega ', t)
\nonumber\\&\hspace{15ex}
        + R_k ^{({\rm o})*} e^{i \omega ' (t-t_0)}
        \delta (\omega - \omega '),
\\
\label{3.31}
&
G_{k\rm{cav}}
         (\omega, \omega ', t) =
        T_k ^{({\rm o})*} A_{k \mathrm{cav}}^*
        \upsilon_k (\omega, \omega ', t),
\\
\label{3.33}
&
G_{k\pm}
        (\omega, \omega ', t) =
        T_k ^{({\rm o})*} A^*_{k\pm}
        \upsilon _k (\omega, \omega ', t)
\nonumber\\&\hspace{15ex}
        + A^{({\rm o})*}_{k\pm} e^{i \omega ' (t-t_0)}
        \delta (\omega - \omega '),
\end{align}
with
\begin{align}
\label{3.35}
&        \upsilon _k(\omega, \omega ', t)  =
        \frac {1}{2\pi}\frac{c}{2n_1 ^* l}
        \frac {e^{-i\omega \Delta t}}{
        \omega - \Omega _k ^* }
\nonumber\\[.5ex]&\hspace{4ex}
\times\,
        \left[
        \frac {e^{ i\omega' (t+ \Delta t -t_0)}
        - e^{i \Omega _k ^*
        (t+ \Delta t-t_0)}}
                {
        \omega' - \Omega _k ^*
        }
\right.
\nonumber\\[.5ex]&\hspace{10ex}
\left.
        -\,
        \frac {e^{i\omega (t+ \Delta t-t_0)}
        - e^{ i\omega' (t+ \Delta t-t_0)}}
                {
        \omega - \omega '
                }
\right].
\end{align}

%%%%%%%%%%%%%%%%%%%%%%%%%%%%%%%%%%%%%%%%%%%%%%%%%%%%%%%%%%%%%%%

\subsection{Nonmonochromatic modes}
\label{sec6.0}

To calculate the
quantum state of the outgoing field,
it is convenient to introduce a unitary,
explicitly time-dependent transformation according to
\begin{align}
  \label{6.4.2}
&      \hat{b}_{
      k
      \mathrm {out}} (\omega , t)
      =
      \sum_i
      \phi _i^*
     (\omega
     ,t
      )
       \hat{b}_{k\mathrm {out}}
       ^{(i)} (t) ,
\\
  \label{6.4.1}
&      \hat{b}_{k\mathrm {out}}^{(i)} (t)
      = \int_
      {(\Delta_k)}
        \!\D\omega\,
      \phi _i(\omega
      ,t
       )
       \hat{b}_{
       k
       \mathrm {out}} (\omega, t),
\end{align}
where, for chosen $t$,
the nonmonochromatic mode functions
$\phi _i (\omega,t) $ are a complete set of
square integrable orthonormal functions:
\begin{align}
  \label{6.2.1}
&     \int _
     {(\Delta_k)}
     \D\omega\,
     \phi _i (\omega
     ,t
     )
     \phi _j^* (\omega
     ,t
     )
     = \delta _{ij} ,
\\
  \label{6.2.2}
&     \sum _i
     \phi _i (\omega
     ,t
     )
     \phi _i^* (\omega '
     ,t
     )
     = \delta (\omega - \omega ') .
\end{align}
Needless to say that the commutation relation
\begin{equation}
  \label{6.6}
  \bigl[\hat{b}
_{k\mathrm{out}}
  ^{(i)}  ( t),
  \hat{b}
_{k\mathrm{out}}
  ^{(j)\dagger} ( t)\bigr]
= \delta_{ij} 
\end{equation}
holds.

Let $\hat{b}_{k\mathrm{out}}^{(1)}(t)$ be the operator
attributed to the
outgoing mode that is related to the
cavity mode through the input-output relation (\ref{3.23}):
\begin{equation}
  \label{6.8}
   \phi_1  (\omega
   ,t
   ) = \frac
          {F_k(\omega , t)}
          {\sqrt{\eta _k(t)}}\,,
\end{equation}
\begin{equation}
 \label{6.69}
         \eta_k(t)  = \int _
         {(\Delta_k)}
         \D\omega\, |F_k(\omega ,t)| ^2.
 \end{equation}
By using Eqs.~(\ref{3.23}), (\ref{6.8}), and
(\ref{6.69}) we may rewrite
(for chosen $k$) Eq.~(\ref{6.4.1}) as
\begin{equation}
\label{6.10}
\hat{b}_{k\mathrm {out}}^{(i)} (t) =
\left\{
\begin{array}{ll}
     \sqrt{\eta_k (t)}\,\hat{a} _k( t _0)+\hat{B}_k ^{(i)} (t)
     &\ \mathrm{if}\ i = 1,\\[2ex]
     \hat{B}_k ^{(i)} (t)
     &\ \mathrm{otherwise},
\end{array}
\right.
\end{equation}
where
\begin{equation}
\label{6.12}
\hat{B}_k^{(i)}(t) = \int_{(\Delta_k)} \D\omega\,
\phi_i(\omega,t)\hat{B}_k(\omega,t).
\end{equation}

Inserting Eq.~(\ref{3.27}) in Eq.~(\ref{6.12}), we may
rewrite $\hat{B}_k^{(i)}(t)$ as
\begin{equation}
\label{6.12-2} \hat{B}_k^{(i)}(t) =
\sqrt{\zeta^{(i)}_{k\mathrm{in}}(t)}\,
\hat{b}^{(i)}_{k\mathrm{in}}(t) +\sum_\lambda
\sqrt{\zeta^{(i)}_{k\lambda}(t)}\, \hat{c}^{(i)}_{k\lambda}(t).
\end{equation}
Here the functions $\zeta^{(i)}_{k\sigma} (t)$ ($\sigma$ $\! =$
$\!\mathrm{in}, \lambda$) read
\begin{equation}
    \label{6.81}
        \zeta^{(i)}_{k\sigma} (t) =
         \int_
        {(\Delta_k)}
        \D \omega \,
        |\chi^{(i)}_{k\sigma} (\omega, t)|^2,
\end{equation}
where
\begin{equation}
\label{6.79}
   \chi^{(i)}_{k\sigma} (\omega, t ) =
    \int_
    {(\Delta_k)}
   \D \omega'\,\phi_i(\omega',t)
   G  _{k\sigma}^* (\omega', \omega , t)
   .
\end{equation}
 The
operators $\hat{b}^{(i)}_{k\mathrm {in}}(t)$ and
$\hat{c}^{(i)}_{k\lambda} (t)$ are defined by
\begin{align}
  \label{6.14.1}
&    \hat{b}^{(i)}_{k\mathrm {in}}(t)
     = \int_
      {(\Delta_k)}
      \D\omega\,
      \frac{\chi^{(i)}_{k\mathrm {in}} (\omega , t )}
      {\sqrt{ \zeta^{(i)}_{k\mathrm {in}} (t)}}\,
       \hat{b}_{k\mathrm {in}} (\omega, t_0),
\\[.5ex]
  \label{6.14.3}
&    \hat{c}^{(i)}_{k\lambda} (t)
     = \int_
      {(\Delta_k)}
      \D\omega\,
      \frac{\chi^{(i)}_{k\lambda} (\omega , t )}
      {\sqrt{ \zeta^{(i)}_{k\lambda} (t)}}\,
       \hat{c}_{k\lambda}(\omega, t_0).
\end{align}

%%%%%%%%%%%%%%%%%%%%%%%%%%%%%%%%%%%%%%%%%%%%%%%%%%%%%%%%%%%%%%%
\subsection{Phase-space functions}
\label{sec6.1}
%%%%%%%%%%%%%%%%%%%%%%%%%%%%%%%%%%%%%%%%%%%%%%%%%%%%%%%%%%%%%%%
Introducing in the characteristic functional
\begin{align}
\label{6.1}
&        C_{k{\rm out}}[\beta(\omega),t]
\nonumber\\[.5ex]&\hspace{2ex}
         =
         \left\langle
         \exp \left[
         \int_
         {(\Delta_k)}
        \!\D\omega\,
         \beta(\omega) \hat{b}^{\dagger}_{
        k
        \mathrm {out}}(\omega, t)
         - \mathrm{H.c.}\right]
        \right\rangle
\end{align}
the operators
$\hat{b}^{(i)}_{k\mathrm{out}}(t)$ according to
Eq.~(\ref{6.4.2})
and taking into account the commutation relation
(\ref{6.6})
we see that the operator exponential factorizes as
\begin{align}
\label{6.1-1}
&\exp\! \left[
         \int_
        {(\Delta_k)}
        \D\omega\,
         \beta(\omega) \hat{b}^{\dagger}_{k\mathrm {out}}(\omega, t)
         - \mathrm{H.c.}\right]
\nonumber\\&\hspace{10ex}
= \prod_i \exp\!\left[
         \beta_i \hat{b}^{(i)\dagger} _{k\mathrm {out}}( t)
         - \mathrm{H.c.}
         \right],
\end{align}
where
\begin{equation}
  \label{6.1.3}
   \beta_i =
   \beta_i(t)
   =
   \int
   _{(\Delta_k)}
    \D\omega\,
\phi_i(\omega,t)
          \beta (\omega ) .
\end{equation}

Let us further consider the case, when the nonmonochromatic modes of
the incoming field
and dissipative channels corresponding to $\hat{B}_k^{(i)}(t)$$,$
$i \neq 1$, are in vacuum state at the initial time $t_0$. Then we may
assume that the resulting characteristic function
factorizes as well, with
\begin{equation}
\label{6.1.1}
        C
    _{k{\rm out}}(\beta_1, t)
         =
         \left\langle
         \exp\!\left[
         \beta_1 \hat{b}^{(1)\dagger} _{k\mathrm {out}}( t)
         -
         \mathrm{H.c.}
         \right]
         \right\rangle
\end{equation}
being the characteristic function of the relevant outgoing mode.
Using Eq.~(\ref{6.10}) and noting that the commutation relation
\begin{equation}
  \label{6.16}
   \bigl[\hat{ a}_k  ( t _0),
   \hat{B}_k^{(1)\dagger} (t)\bigr] = 0
\quad(t \ge t_0)
\end{equation}
holds (see Appendix \ref{app.2.6}), we may rewrite
Eq.~(\ref{6.1.1}) as
\begin{align}
\label{6.1.2}
&        C
    _{k{\rm out}}(\beta_1, t)
         = \left\langle
         \exp\!\left[\beta_1\sqrt{\eta_k(t)}
         \hat{a}_k^\dagger(t_0) - \mathrm{H.c.}
         \right]\right.
\nonumber\\&\hspace{17ex}\times\,
         \left.\exp\!\left[\beta_1
         \hat{B}_k^{(1)\dagger}(t) - \mathrm{H.c.}
         \right]
         \right\rangle.
\end{align}
Noting that
according to
Eq.~(\ref{6.12-2})
$\hat{B}_k^{(1)\dagger}(t)$ is a functional of
$\hat{b}_{k\mathrm {in}} (\omega, t_0)$ and
$\hat{c}_{k\lambda} (\omega, t_0)$ and assuming that the
density operator (at the initial time $t_0$) factorizes
with respect to the cavity field, the incoming field, the
dissipative channels, we obtain
\begin{align}
\label{6.1.2-1}
&        C
    _{k{\rm out}}(\beta_1, t)
         = \left\langle
         \exp\!\left[\beta_1\sqrt{\eta_k(t)}
         \hat{a}_k^\dagger(t_0) - \mathrm{H.c.}
         \right]\right\rangle
\nonumber\\&\hspace{17ex}\times\,
         \left\langle\exp\!\left[\beta_1
         \hat{B}_k^{(1)\dagger}(t) - \mathrm{H.c.}
         \right]
         \right\rangle.
\end{align}

Making use of the commutation relations (\ref{5.9}) and
\begin{equation}
  \label{6.14}
   \bigl[
    \hat{B}_k ^{(1)} (t),\hat{B}_k ^{(1) \dagger} (t)
   \bigr] = 1- \eta_k (t),
\end{equation}
which follows from Eq.~(\ref{6.10}) together with the
commutation relations (\ref{5.9}), (\ref{6.6}), and (\ref{6.16}),
the further evaluation of
Eq.~(\ref{6.1.2-1}) is straightforward.
Following
Ref.~\cite{khanbekyan:043807} and calculating the
characteristic function $C_\mathrm {out}(\beta,t;s)$
$\!\equiv$ $\!C_{k\mathrm {out}}(\beta_1,t;s)$ in
$s$ order of the quantum state of the relevant outgoing field,
we may express it in terms of the characteristic
function $C(\beta';s')$ $\!\equiv$
$\!C_k(\beta';s')$
of the quantum state of the initially excited cavity mode
and the characteristic functions
$C_\sigma(\beta_\sigma;s_\sigma)$ $\!\equiv$
$\!C_{k\sigma}(\beta_\sigma;s_\sigma)$
of the quantum states of the
incoming field ($\sigma$ $\!=$ $\!\mathrm{in}$)
and the dissipative channels ($\sigma$ $\!=$ $\!\lambda$) as
\begin{align}
\label{6.47}
&        C_\mathrm {out}(\beta,t;s)
        =
        e^{ - {\textstyle\frac{1}{2}}
          \xi(t)
           |\beta|^2}
             C
        \bigl[\sqrt {\eta(t)}\,\beta; s'\bigr]
\nonumber\\&\hspace{18ex}\times\,
    \prod _{\sigma }\!
        C
        _{\sigma} \!
       \Bigl[
       \sqrt{\zeta _{\sigma } (t) }
       \,\beta ; s_
{\sigma}
       \Bigr]
        ,
\end{align}
where
\begin{equation}
\label{6.77}
   \xi(t)=
        \eta(t)
s'
+
        \sum _{\sigma } \zeta _{\sigma} (t)
   s_{\sigma }
  -s
\end{equation}
[$\eta(t)$ $\!\equiv$ $\!\eta_k(t)$,
$\zeta_\sigma(t)$ $\!\equiv$ $\!\zeta^{(1)}_{k\sigma}(t)$].

{F}rom Eq.~(\ref{6.47}) the phase-space function in $s$ order
can be derived to be
\begin{align}
\label{6.75}
&        P_\mathrm {out}( \alpha,t ; s )
         = \frac {2} {\pi}
         \frac {1}{\xi (t)}
\nonumber\\&\hspace{2ex}\times\,
        \int \D ^2 \alpha ' \,P
        (\alpha '; s')
        \prod_{\sigma }
        \int
        \D ^{2}\alpha_{\sigma }\,
        P _{\sigma }(\alpha  _{\sigma } ; s _{\sigma })
\nonumber\\&\hspace{2ex}\times\,
        \exp\biggl[
          -\frac {2} {\xi(t)}
\Bigl|
\sqrt{
        \eta(t)
        }\, \alpha'
        \!+\!
        \sum_{\sigma}
        \sqrt{
         \zeta _{\sigma }
         (t)}\,
        \alpha _{\sigma }
        \!-\!\alpha
\Bigr|^2
\biggr],
\end{align}
provided that
\begin{equation}
\label{6.77.3}
\xi(t)\ge 0,
\end{equation}
where the equality sign must be understood as a limiting process.
To calculate $\eta(t)$ [Eq.~(\ref{6.69})] and
$\zeta _{\sigma } (t)$ [Eq.~(\ref{6.81})]
we make use of Eqs.~(\ref{3.25}), (\ref{3.29})--(\ref{3.35}),
(\ref{6.8}), and (\ref{6.79}). Straightforward calculation yields
[$T$ $\!=$ $\!T_k$,
$A_{\pm}$ $\!=$ $\!A_{k\pm}$,
$T^{({\rm o})}$ $\!=$ $\!T_k^{({\rm o})}$,
$A^{({\rm o})}_{\pm}$ $\!=$ $\!A^{({\rm o})}_{k\pm}$,
$R^{({\rm o})}$ $\!=$ $\!R_k^{({\rm o})}$,
$\gamma_{\sigma}$ $\!=$ $\!\gamma_{k\sigma}$,
$\Gamma$ $\!=$ $\!\Gamma_{k}$]
\begin{align}
\label{6.87} &
\eta(t\rightarrow \infty) = \frac
{\gamma_\mathrm{rad}^\mathrm{(o)} }
        {
    \Gamma
        }
,
\\
\nonumber
&   \zeta _\mathrm{in}
   (t\rightarrow \infty) =
          \frac{\gamma _\mathrm{rad}^\mathrm{(o)}\gamma _\mathrm{rad}}
             {
            \Gamma
              ^2}
        +
 \abq{ R ^\mathrm{(o)}}
\\&
    \label{6.91}
    \hspace{2ex}
    +\,
    \frac{c} {2n_1^*l}
        \frac{R^\mathrm{(o)} T^* T^{\mathrm{(o)}*}}
            {\Gamma}
     +
    \frac{c} {2n_1l}
        \frac{R^{\mathrm{(o)}*} T T^\mathrm{(o)}}
            {\Gamma}
              \,,
\\
    \nonumber
&   \zeta_{\pm}
   (t\rightarrow \infty) =
   \frac {
          \gamma _\mathrm{rad}^\mathrm{(o)}
            \gamma _{\pm}
          }
          {\Gamma ^2}
          +
          \abq{ A_{\pm} ^\mathrm{(o)}}
\\&
    \label{6.91.3}
    \hspace{2ex}
    +\,
    \frac{c} {2n_1^*l}
        \frac{A^\mathrm{(o)}_{\pm} A^*_{\pm} T^{\mathrm{(o)}*}}
            {\Gamma}
     +
    \frac{c} {2n_1l}
        \frac{A^{\mathrm{(o)}*}_{\pm} A_{\pm} T^\mathrm{(o)}}
            {\Gamma}
              \,,
\\
   \label{6.91.4}
&   \zeta _\mathrm{cav}
   (t\rightarrow \infty) =
 \frac{ \gamma _\mathrm{rad}^\mathrm{(o)}\gamma _\mathrm{cav}}
        {
    \Gamma
        ^2}\,,
\end{align}
where the damping rates $\gamma_\mathrm{rad}$, $\gamma
_\mathrm{abs}$, and $\gamma _\lambda$ are defined according to
Eqs.~(\ref{5.3}) and Eq.~(\ref{5.7}), and
\begin{equation}
\label{6.91.5}
\gamma_{\mathrm{rad}}^\mathrm{(o)}
\equiv
\gamma_{k\mathrm{rad}}^\mathrm{(o)}
     = \frac{c}{2 |n_1| l}   |T^\mathrm{(o)} _k|^2 .
\end{equation}
Eq.~(\ref{6.75}) together with Eqs.~(\ref{6.87})--(\ref{6.91.4})
generalizes the result in Ref.~\cite{khanbekyan:043807} 
since
it fully takes into account the noise associated with the
dissipative channels.

%%%%%%%%%%%%%%%%%%%%%%%%%%%%%%%%%%%%%%%%%%%%%%%%%%%%%%%%%%%%%%%
\subsection{Thermal noise}

Let us consider the typical case of
the dissipative channels being in thermal states, i.e.,
\begin{equation}
   \label{6.99}
        W
        _{
        \lambda
        }
        (\alpha
       )
        =
        \frac {2} {\pi}
        \frac {1} {1+ 2\bar{n}_ {
    \lambda
    }  }\,
        e^{-
  %   {\textstyle
%         \frac {2|\alpha
%      |^2} {1+ 2\bar{n}_ {\lambda}}
%           }
   2|\alpha |^2/(1+ 2\bar{n}_ {\lambda})      
        }
\end{equation}
($\bar{n}_ {\lambda}$, average number of thermal quanta)
and calculate the Wigner function of the quantum state of the
relevant
outgoing mode.
Inserting Eq.~(\ref{6.99}) into Eq.~(\ref{6.75}),
after having set $s_{\lambda}$ $\!=$ $\!0$ therein,
performing the $\alpha_\lambda$
integrations, and
setting
$s$ $\!=$ $\!s '$ $\!=$ $s
%^{(1,2)}
_{\mathrm{in}}$$=$ $\!0$, we derive
\begin{align}
\label{6.109}
&    W_{{\rm out}}(\alpha, t)
        = \frac {2} {\pi}
\frac{1} {\xi^W (t)}
\nonumber
\\&\hspace{2ex}\times\,
     \int \D ^2 \alpha'
     \int \D ^2 \beta\,
      W(\alpha') W_{{\rm in}}(\beta)
 \nonumber
 \\&\hspace{4ex}\times\,
        \exp\!\left[\! -\frac {2|\sqrt{
        \eta(t)
        }\, \alpha'
    +
     \sqrt{
         \zeta _{\mathrm{in} }
         (t)}\,
        \beta
          -\alpha |^2 }
    {\xi ^W (t)}
        \right]
     ,
\end{align}
where
\begin{equation}
    \label{6.131}
        \xi ^W(t)
            =
             1-
        \eta(t)
        -
        \zeta
        _{\mathrm{in} }(t)
        + 2
       \sum_ {\lambda}
       \bar{n} _{\lambda} \zeta
       _{\lambda}(t) .
\end{equation}

Let us consider Eq.~(\ref{6.109}) together with Eq.~(\ref{6.131})
for some typical situations in more detail.\\
(i) When the incoming field is in the vacuum state (unused input
port, $W_\mathrm{in}(\beta)$ $\!=$ $\!2\pi^{-1}e^{-2|\beta|^2}$)
and the dissipative channels---in particular, the coupling
mirror---are in the vacuum state as well, then almost perfect
extraction of the quantum state of the cavity mode requires the
condition
\begin{equation}
\label{6.109.4}
\frac{\eta(t)}{1-\eta(t)} \gg 1
\end{equation}
to be satisfied. In other words, on recalling Eq.~(\ref{6.87}),
the nonradiative cavity-field decay rate must be small compared
with the radiative one,
\mbox{$\gamma_\mathrm{abs}/\gamma_\mathrm{rad}^\mathrm{(o)}$ $\!\ll$ $\!1$}---a
condition that can be hardly satisfied for a high-$Q$ cavity
presently
\cite{rempe:363, hood:033804}. How small---depends on the
nonclassical features of the quantum field to be extracted.
Notice, this condition can be
also
obtained by means of quantum stochastic approach
to the problem
\cite{khanbekyan:043807}.\\
(ii) When the input port is unused but the dissipative channels
are thermally excited, then,
as one can easily see from Eq.~(\ref{6.109}), the condition to
ensure nearly perfect extraction of the
quantum state of the
cavity field is
\begin{equation}
    \label{6.110}
        \frac{
        \eta(t)
        }{
        1-
        \eta(t)
        + 2
        \sum_ {\lambda}
        \bar{n} _{\lambda} \zeta _{\lambda}
        } \gg 1
        \, .
\end{equation}
The condition Eq.~(\ref{6.110}) strengthens even more the requirement
of smallness of nonradiative cavity-field decay
rate
compared
with the radiative
one.
Particularly, the value of $\sum_
{\lambda} \bar{n}_{\lambda} \zeta _{\lambda}(t)$ should be as small
as possible to ensure that the effect of thermal noise effectively
does not play a role. This is obviously the case when both
$\bar{n} _{\lambda}$ and $\zeta_{\lambda}(t)$ are sufficiently small. Needless
to say that small values of $\bar{n}_{\lambda}$ require sufficiently low
temperatures.
For
cavities with high-quality
mirrors \cite{rempe:363, hood:033804} with the finesse of
several hundred thousands,
the second, the third and the forth terms on the right-hand side of  
Eq.~(\ref{6.91.3})
are of an order smaller magnitude than
the first term on the right-hand side of  
Eq.~(\ref{6.91.3}), as well as
$\zeta _\mathrm{cav}$,
Eq.~(\ref{6.91.4}),
and may be therefore
disregarded in the
sum
$\sum_{\lambda} \bar{n}_{\lambda} \zeta _{\lambda}(t)$.
That is to say, in case
of unused input port
dissipation due to
absorption in the coupling mirror can be
effectively
described by adding
appropriate Langevin noise forces in Eq.(\ref{2.71}).\\
(iii) To describe typical problems on engineering of nonclassical
states of light let us assume, that the dissipative channels
are again thermally excited
and
the incoming field
mode with the
mode
function
$\! \chi^{(1)}_{k \mathrm{in}} (\omega, t )$
according to
Eq.~(\ref{6.79}) is
prepared
in some
nonclassical state. Then, the quantum state of the outgoing
field mode is
the one
of the cavity-mode
superposed with
the
reflected
incoming field
mode
as well as the
modes of the (thermally excited) dissipative channels.
The weights of the modes of the  incoming field and the cavity-mode field
in the resulting superposition
are defined respectively by the fractions
\begin{align}
    \label{6.114}
&
        \frac{
        \zeta
       _{\mathrm{in}}(t)
        }{
             1-
        \eta(t)
        -
        \zeta
        _{\mathrm{in} }(t)
        + 2
       \sum_ {\lambda}
       \bar{n} _{\lambda} \zeta
       _{\lambda}(t)
        } 
        \,  ,
\\ \label{6.116}
&
 \frac{
        \eta
        (t)
        }{
             1-
        \eta(t)
        -
        \zeta
        _{\mathrm{in} }(t)
        + 2
       \sum_ {\lambda}
       \bar{n} _{\lambda} \zeta
       _{\lambda}(t)
        } 
        \,
.
\end{align}
Notice, that dropping the absorption in the coupling mirror,
\mbox{$|R_k^\mathrm{(o)}|$ $\!=$ $\!1$}, and Eq~(\ref{6.91}) reduces
to \mbox{$\zeta _\mathrm{in}(t\rightarrow \infty)$ $\!=$
  $[1-\eta(t\rightarrow \infty)]^2$}. The additional noise associated
with the coupling mirror reduces the fraction of the input field in
the resulting superposition, and,
therefore, represents the absorption of the incoming field mode in the
coupling mirror.
For high-$Q$ cavities with the finesse of
several hundred thousands \cite{rempe:363, hood:033804}, the
unwanted losses in the coupling mirror
%described by the last two terms in Eq.~(\ref{3.11}) 
reduce the weight of the incoming field mode by about 
$50\%$.
In this way, the quantum state of the output mode carries additional
noise.

Notice,
to obtain the above given results we have
assumed, that the  nonmonochromatic modes of the incoming field and
the 
dissipative channels corresponding to $\hat{B}_k^{(i)}(t)$$,$
$i \neq 1$,
are initially
prepared in the vacuum state. In practice, this is not
necessarily the case, especially with regard to the dissipative channels
associated with the coupling mirror, due to the finite number of
thermal quanta and the impossibility to prepare the mode of a
dissipative channel. 
As a consequence, additional noise is fed into the cavity.
 Moreover, 
it is straightforward to prove with the use of Eqs.~(\ref{5.51}), (\ref{5.53})
and (\ref{6.16}), that
there are 
necessarily more than one
(nonmonochromatic) mode functions of 
 the outgoing field,
including the one 
corresponding to $\hat{b}^{(1)}_{k\mathrm {out}}(t)$,
that 
lie 
in the relevant frequency interval, defined by the bandwidth $\Gamma
_k$ of the
cavity mode resonance frequency $\omega _k$. 
Therefore,
if
the nonmonochromatic modes of
the dissipative channels 
and the incoming field
corresponding to $\hat{B}_k^{(i)}(t)$$,$
$i \neq 1$
are initially prepared in another than the vacuum state, the quantum
state of the output field in the relevant frequency interval is a
%(in general unfactorized) 
mixture of modes
of the outgoing field corresponding to 
$\hat{b}^{(i)}_{k\mathrm  {out}}(t)$
 including the relevant one, which corresponds to
$\hat{b}^{(1)}_{k\mathrm {out}}(t)$.  
The mode analysis for this case will be performed in detail in a
forthcoming paper.

%%%%%%%%%%%%%%%%%%%%%%%%%%%%%%%%%%%%%%%%%%%%%%%%%%%%%%%%%%%%%%%
\section{Summary and Conclusions}
\label{sec9}
%%%%%%%%%%%%%%%%%%%%%%%%%%%%%%%%%%%%%%%%%%%%%%%%%%%%%%%%%%%%%%%

Within the frame of exact quantum electrodynamics in causal media
we have studied the input-output problem of a high-$Q$ cavity.
Making use of the representation of the quantized electromagnetic
field in dispersing and absorbing planar (dielectric) multilayers
as given in Ref.~\cite{khanbekyan:063812}, we have considered a
one-dimensional cavity bounded by a perfectly reflecting mirror
and a fractionally transparent mirror, which is responsible for
the input-output coupling. In order to study the effect of
unwanted losses such as absorption losses, we have allowed both
the medium inside the cavity and the coupling mirror to be
absorbing, by attributing to them complex permittivities.
Moreover, we have assumed that there are also active atoms inside
the cavity, which are supposed to interact with the
medium-assisted electromagnetic field via electric-dipole
coupling.

We have calculated the electromagnetic field both inside and
outside the cavity. It has turned out that in a coarse-grained
approximation, i.e., on a time scale that is large compared with
the inverse separation of two neighboring cavity resonance
frequencies, the intracavity field may be expressed in terms of
standing waves, and bosonic operators associated with them can be
introduced which obey quantum Langevin equations. In this
approximation, the radiative losses due to the input-output
coupling and the absorption losses can be regarded as representing
independent dissipative channels, each giving rise to a damping
rate and a corresponding Langevin noise force. The result shows
that the Hamiltonian used in quantum noise theories
\cite{gardiner:3761} to treat a leaky cavity can be simply
complemented by bilinear interaction energies between the cavity
modes and appropriately chosen dissipative channels to model
unwanted losses such as absorption losses.

However, this intuitive concept fails with respect to the operator
input-output relations in general. As we have shown, the absorption
losses attributed to the coupling mirror give rise to additional
force terms in the input-output relations which cannot be simply inferred
from the above mentioned interaction energies between the cavity
modes and the dissipative channels introduced to model the
mirror-assisted absorption. Hence the input-output relations
obtainable from standard quantum noise theories would be incomplete.

Finally we have used the exact operator input-output relations to
explicitly calculate the quantum state of the outgoing field as a
function of time, assuming that the quantum state of the cavity
field is known at some initial time. To be more specific, we have
restricted our attention to a single cavity mode and assumed that
the process of quantum state preparation is sufficiently short
compared with the decay time of the mode under consideration, so
that the time scales of quantum state preparation and extraction
from the cavity are well separated from each other. Introducing
the relevant modes of the incoming and outgoing fields, i.e., the
modes the cavity mode couples to, we have expressed the
$s$-parameterized phase-space function of the quantum state of the
relevant outgoing mode in terms of the phase-space functions of
the quantum states of the cavity mode, the relevant incoming mode,
and the dissipative degrees of freedom responsible for
unwanted losses.
It should be mentioned that the generalization to
more than one cavity mode initially excited
is straightforward.

%%%%%%%%%%%%%%%%%%%%%%%%%%%%%%%%%%%%%%%%%%%%%%%%%%%%%%%%%%%%%%

\begin{acknowledgments}
This work was supported by the Deutsche Forschungsgemeinschaft.
A.A.S. and W.V. gratefully acknowledge support by the Deutscher
Akademischer Austauschdienst.
\end{acknowledgments}

\appendix
%%%%%%%%%%%%%%%%%%%%%%%%%%%%%%%%%%%%%%%%%%%%%%%%%%%%%%%%%%%%%%%
\section{\label{app.1}Recursion formulas for Fresnel coefficients}
%%%%%%%%%%%%%%%%%%%%%%%%%%%%%%%%%%%%%%%%%%%%%%%%%%%%%%%%%%%%%%%

To calculate
$r_{ij}$ and $t_{ij}$, we
first note, that
in the case
$|i$ $\!-$ $\!j|$ $\!=$ $\!1$,
i.e., single-interface transmission,
they are defined according to
\begin{align}
      \label{a1.1}
&      r_{ij} \equiv r_{i/j} =
      \frac{\beta_i-\beta_j}{\beta_i+\beta_j}
      = -r_{j/i} ,
\\
      \label{a1.3}
&      t_{ij} \equiv t_{i/j} = (1+r_{i/j})
      = \frac{\beta_i}{\beta_j}\,t_{ji},
\end{align}
leading to
\begin{equation}
      \label{a1.5}
      t_{ij}t_{ji}-r_{ij}r_{ji} = 1.
      \end{equation}
In the general case, the
relations
(see Ref.~\cite{tomas:052103})
\begin{align}
      \label{a1.7}
&      r_{i/j/k} =
    \frac{r_{i/j}+(t_{i/j}t_{j/i}-r_{i/j}r_{j/i})r_{j/k}
      e^{2i\beta_jd_j}}
    {  1-r_{j/i}r_{j/k} e^{2i\beta_jd_j}}
\,,
\\
      \label{a1.9}
&      t_{i/j/k} = \frac{t_{i/j}t_{j/k}
      e^{i\beta_jd_j}}
    { 1-r_{j/i}r_{j/k} e^{2i\beta_jd_j}}
\end{align}
[$\mbox{min}(i,k)$ $\!\leq$ $\!j$ $\!\leq$
$\!\mbox{max}(i,k)$]
hold.
With these formulas at hand, we can calculate recursively all the
quantities $r_{i/j}$ and $t_{i/j}$, since
\begin{align}
\label{a1.13}
&      r_{ik} \equiv
      r_{i/k} = r_{i/j/k}
,
\\
\label{a1.13-1}
&      t_{ik} \equiv
      t_{i/k} = t_{i/j/k}
\end{align}
for any $j$ with
$\mbox{min}(i,k)$ $\!\leq$ $\!j$ $\!\leq$ $\!\mbox{max}(i,k)$.
Note that in the system under consideration
\mbox{$r_{10}$ $\!=$ $\!-1$}, since
perfect reflection from the left-side mirror
of the cavity field has been postulated.

%%%%%%%%%%%%%%%%%%%%%%%%%%%%%%%%%%%%%%%%%%%%%%%%%%%%%%%%%%%%%%%
%\subsection
\section{\label{app.1.3}Derivation of Eq.~(\ref{3.3})}

Inserting
Eq.~(\ref{1.49}) in
Eq.~(\ref{1.45-1}) for $j$ $\!=$ $\!3$, we may write
$\underline{\hat{ E}}_{ {\rm out}, {\rm free}} ^{(3)}(z, \omega, t)$
at $z$ $\!=$ $\!0^+$ (cf Fig.~\ref{fig}) as
\begin{align}
      \label{a1.16}
&    \underline{\hat{ E}}_{ {\rm out}, {\rm free}}
     ^{(3)}
     (z, \omega, t)\bigr|_{z=0^+}
\nonumber\\&
    = \frac { t_{13}e^{i \beta_1 l}}
            {D_1 }
    \left[ \hat {C} ^{(1)}_{<+}(l, \omega, t)
        - \hat {C} ^{(1)}_{<-}(l, \omega, t)\right]
\nonumber\\&
\hspace{1ex}
       + \frac {t_{23}e^{i \beta_2 d} } { D_2}
    \!\left[ \hat {C} ^{(2)}_{+}(\omega, t )
       + r_{20} \hat {C} ^{(2)}_{-}(\omega, t ) \right]
       \!+ r_{30}\hat {C} ^{(3)}_{-}(\omega, t ),
\end{align}
where $\hat {C} ^{(1)}_{<\pm}(l, \omega, t )$,
$\hat {C} ^{(2)}_{\pm}(\omega, t )$,
and $\hat {C} ^{(3)}_{-}(\omega, t )$ are given by
Eqs.~(\ref{2.5}), (\ref{2.9}),
and (\ref{2.11}), respectively.
Using Eqs.~(\ref{1.57}), (\ref{2.3}), and (\ref{a1.7}),
the following relations can be easily proved:
\begin{align}
      \label{a1.17}
&        \frac {1}{D_2}
        \left(1 \pm r_{20}
        e^{i \beta _2 d }\right)
        -
        \frac {1}{D_2 '}
        \left(1 \pm r_{21}
        e^{i\beta _2 d }\right)
\nonumber\\&\hspace{9ex}
        =
        \mp
        \frac {t_{13}}{D_1 D_2'}
        \frac{t_{21}}{t_{23}}
        \left(1 \pm r_{23}
        e^{i \beta _2 d}\right)
        e^{2i \beta _1 l} ,
\\
      \label{a1.19}
&        r_{30} - r_{31}
        =
        -
        \frac {t_{13} t_{31}}{D_1 }\,
        e^{2i \beta _1 l}.
\end{align}
Combining Eq.~(\ref{a1.16}) with Eqs.~(\ref{a1.17})
and (\ref{a1.19}), we arrive at Eq.~(\ref{3.3}).

%%%%%%%%%%%%%%%%%%%%%%%%%%%%%%%%%%%%%%%%%%%%%%%%%%%%%%%%%%%%%%%%
\section{\label{app.2.1}Derivation of Eqs.~(\ref{5.1})--(\ref{5.7})}

We solve Eq.~(\ref{2.15}) [equivalently, Eqs.~(\ref{2.19.3})
and (\ref{2.19.5})] by iteration to obtain $\Gamma_k$ in leading
order as
\begin{equation}
  \label{a2.9}
  \Gamma_k  = \frac{c}{2 n_1 l}
 (1-|r_{13}|^2),
\end{equation}
where $n_1$, $r_{13}$, and the parameter introduced in the following
are taken at the (unperturbed) frequency
$\omega_k$. By means of
\begin{equation}
\label{a1.a6}
    r_{13}
    = \frac {-r_{21} + r_{23}e^{2 i \beta _2 d_2}}
    {1-r_{23} r_{21} e^{2 i \beta _2 d_2} }
\end{equation}
Eq.~(\ref{a2.9}) can be rewritten as
\begin{align}
  \label{a2.9.1}
  \Gamma_k  &= \frac{1}{|D_2' |^2} \frac {4 n_1} {|n_1+ n_2|^2}
        \left[
            n_2 ' \left(1 - |r_{23}|^2 e^{-4 \beta _2 '' d}\right)
        \right.
\nonumber \\&\quad
        \left.
        +\,  i n_2 '' \left(r_{23} ^* e^{- 2 i \beta _2^* d } -
        r_{23}  e^{ 2 i \beta _2 d }
        \right)
\right] .
\end{align}
Further, from Eqs.~(\ref{2.61}) ($\omega$ $\!=$ $\!\omega_k$)
and (\ref{a1.9}) we find
\begin{equation}
  \label{a2.31}
    |T_k|^2 = \frac
        {16 n_1 |n_2|^2n_3 ' }
            {|D_2 '|^2 |n_1 + n_2 | ^2 |n_2 + n_3 | ^2 }
            \,
            e^{-2 \beta _2 '' d}
                .
\end{equation}
Making use of Eqs.~(\ref{2.57}) and (\ref{2.59}), we derive
($\omega$ $\!=$ $\!\omega_k$)
\begin{align}
  \label{a2.33}
&\hspace{-2ex}
        \sum_\lambda |A_{k\lambda}|^2
        = \frac
        {4 n_1 }
            {|D_2 '|^2 |n_1 + n_2 | ^2 }
            \,
            e^{-\beta _2 '' d}
\nonumber \\&\hspace{-1ex}\times\,
        \left[
            n_2 ' \left(e^{\beta _ 2'' d }
            - e^{-\beta _2 '' d}\right)
            \left(1+ |r_{23}|^2 e^{-2\beta _2 '' d}\right)
        \right.
\nonumber \\&
    \left.
    -\,i n_2 ''\left(e^{i\beta _ 2' d } - e^{-i\beta _2 ' d}\right)
     \left (r_{23}e^{i\beta _2  d}  + r_{23}^*e^{-i\beta _2^* d}\right)
    \right].
\end{align}
Thus, combining Eqs.~(\ref{a2.9.1})--(\ref{a2.33}), we arrive at
\begin{equation}
\label{a2.3}
        \Gamma_k  =
        \frac{c}{2 |n_1| l}\,\Bigl(
          |T_k|^2 +
        \sum_\lambda |A_{k\lambda}|^2
        \Bigr)
,
\end{equation}
which matches Eq.~(\ref{5.1}) together with
Eqs.~(\ref{5.3}) and (\ref{5.7}).

\section{\label{app.2.3}Proof of the commutation relation (\ref{5.9})}
%%%%%%%%%%%%%%%%%%%%%%%%%%%%%%%%%%%%%%%%%%%%%%%%%%%%%%%%%%%%%%%

To prove the commutation relation (\ref{5.9}), we recall the
definition of $\hat{E}_{k }^{(1)}(z, t)$, namely
\begin{equation}
      \label{a4.1}
      \hat{E}_{k }^{(1)}(z, t)
      = \int_{(\Delta_k)} \D\omega\,
      \underline{\hat{E}} ^{(1)}(z,\omega,t),
\end{equation}
where $\underline{\hat{E}} ^{(1)}(z,\omega,t)$ is defined by
Eq.~(\ref{1.45-1}) for $j$ $\!=$ $\!1$.
Using Eq.~(\ref{1.47}), employing the integral relation
\begin{equation}
  \label{a4.3}
  \mathrm{Im}\,G
  (z_1, z_2,\omega)
  = \frac {\omega ^2} {c^2}
  \int  \D x\,
  \varepsilon'' (x, \omega)
  G(z_1, x, \omega)G^*(z_2, x, \omega),
\end{equation}
and recalling the commutation relation (\ref{1.11}),
after some algebra
we find that
\begin{align}
  \label{a4.5}
&    \bigl[ \hat{E}_{k }^{(1)}(z_1, t)
    ,
 \hat{E}_{k' }^{(1)\dagger}(z_2, t)
    \bigr]
\nonumber \\&\hspace{5ex}
    = \delta_{kk'}
    \int_{(\Delta_k)} \D\omega\,
    \omega ^2\,
    \frac{\mu _0 \hbar}
    {\pi \mathcal{A}}
    \,
    \mathrm{Im}\,G ^{(11)}
  (z_1, z_2,\omega)
    .
\end{align}
To perform the integration we
extend the lower (upper) integration limit to $-\infty$
($\infty$), and rewrite
$ \mathrm{Im}\,G ^{(11)}(z_1, z_2,\omega)$
\mbox{$\!=$
$\![G ^{(11)}(z_1, z_2,\omega)$ $\!-$
$\!G ^{(11)\ast}(z_1, z_2,\omega)]$$/(2i)
$}.
Then, recalling the definition of
$ G ^{(11)}(z_1, z_2,\omega)$
[$G ^{(11)\ast}(z_1, z_2,\omega)$] from Eqs.~(\ref{1.49}) --
(\ref{1.57}) with $j=1$, we evaluate the integral applying the
residue theorem for the poles determined by the zeroes of the function
$D_1(\omega)$ [$D_1^*(\omega)$].
Thus, for sufficiently high-$Q$ cavities,
$\Gamma_{k}$ $\!\ll $ $\!\Delta_k$ we obtain
\begin{align}
  \label{a4.7}
&\hspace{-2ex}
    \bigl[ \hat{E}_{k }^{(1)}(z_1, t)
    ,
 \hat{E}_{k' }^{(1)\dagger}(z_2, t)
    \bigr]
\nonumber \\&\hspace{2ex}
    = \delta_{kk'}
    \frac {\hbar \omega _k}
    {\epsilon _0 |\varepsilon _1| l \mathcal{A} }
    \,
    \sin [\beta _1(\omega _k) z_1]
    \sin [\beta _1^*(\omega _k) z_2]
  .
\end{align}
Comparing Eq.~(\ref{a4.7}) with Eq.~(\ref{2.51})
[together with Eq.~(\ref{2.53})], we then easily see
that the commutation relation (\ref{5.9}) holds.

\section{\label{app.2.2}Proof of the commutation relation (\ref{5.21})}
%%%%%%%%%%%%%%%%%%%%%%%%%%%%%%%%%%%%%%%%%%%%%%%%%%%%%%%%%%%%%%%

{F}rom Eq.~(\ref{3.7}) it follows that
\begin{align}
    \label{a8.1}
&
    \bigl[\hat{ b}_{
k
    \mathrm{out}}  (t),
    \hat{b}_{
k'
     \mathrm{out}}^\dagger (t')\bigr] =
    \frac{1} {2\pi}
    \int
_{(\Delta_k)}
    \mathrm{d}\omega\,
    \int
_{(\Delta_{k'})}
    \mathrm{d}\omega '\,
    \abs{\alpha _{ \mathrm{out}}}^2
\nonumber \\&\hspace{2ex}\times\,
    \frac{\pi\mathcal{A}}{\mu_0c\hbar\sqrt{\omega \omega '}}\,
    \bigl[
      \uh{ E}^{(3)} _{k\mathrm{out}}(0^+, \omega, t)
,
       \uh{ E}^{(3)\dagger} _{k'\mathrm{out}}(0^+, \omega ', t ')
    \bigr]
,
\end{align}
which in the source-quantity representation reads as
[cf. Eqs.~(\ref{2.0}) and (\ref{2.53-1})]
\begin{align}
    \label{a8.3}
&    \bigl[\hat{ b}_{
k
     \mathrm{out}}  (t),
    \hat{b}_{
k'
     \mathrm{out}}^\dagger (t')\bigr]
\nonumber\\&\hspace{2ex}
    = \frac{1} {2\pi}
    \int
_{(\Delta_k)}
    \mathrm{d}\omega
    \int
_{(\Delta_{k'})}
    \mathrm{d}\omega '\,
    \abs{\alpha _{ \mathrm{out}}}^2
    \frac{\pi\mathcal{A}}{\mu_0c\hbar\sqrt{\omega \omega '}}
\nonumber \\&\hspace{4ex}\times
\left\lbrace
    \bigl[
      \uh{ E}^{(3)} _{k{\rm out}, {\rm free}}(0^+\!, \omega, t)
,
       \uh{ E}^{(3)\dagger} _{k'{\rm out}, {\rm free}}(0^+\!, \omega ', t ')
    \bigr]
    \right.
\nonumber \\&\hspace{6ex}
    +
    \bigl[
      \uh{ E}^{(3)} _{
     k{\rm s}}(0^+\!, \omega, t)
,
       \uh{ E}^{(3)\dagger} _{
     k'{\rm s}}(0^+\!, \omega ', t ')
    \bigr]
\nonumber \\&\hspace{6ex}
    +
       \bigl[
      \uh{ E}^{(3)} _{k{\rm out}, {\rm free}}(0^+\!, \omega, t)
,
       \uh{ E}^{(3)\dagger} _{
       k'{\rm s}}(0^+\!, \omega ', t ')
    \bigr]
\nonumber \\&\hspace{6ex}
\left.
    +
    \bigl[
      \uh{ E}^{(3)} _{
      k{\rm s}}(0^+\!, \omega, t)
,
       \uh{ E}^{(3)\dagger} _{k'{\rm out}, {\rm free}}(0^+\!, \omega ', t ')
    \bigr]
\right\rbrace .
\end{align}
Using Eq.~(\ref{1.40}) one easily finds
\begin{align}
    \label{a8.8}
&       \bigl[
      \uh{ E}^{(3)} _{
     k{\rm s}}(0^+\!, \omega _1, t _1)
,
       \uh{ E}^{(3)\dagger} _{
     k{\rm s}}(0^+\!, \omega _2, t _2)
    \bigr]
    =
    \frac{1} {\pi ^2 \epsilon_0 ^2 \mathcal{A} ^2}
    \frac{ \omega_1 ^2 \omega_2 ^2} {c^4}
\nonumber \\[.5ex]&\hspace{1ex}
\times\,
    \sum_{AA'}
    \mathrm{Im}\,G^{(13)}(z_{A'}, 0^+\!, \omega _1)
    \mathrm{Im}\,G^{(13)} (z_{A}, 0^+\!, \omega_2 )
\nonumber \\[.5ex]&\hspace{3ex}
\times\,  \int \D t'\, \int \D t''\,
 \Theta(t_1-t') \Theta(t_2-t'')
\nonumber \\[.5ex]&\hspace{5ex}
\times\,
    e^{-i\omega_1 (t_1-t')}
    e^{i\omega_2 (t_2-t'')}
    \bigl[
    \hat{ d}_{A'} (t'),
    \hat{ d}_{A} (t'')\bigr].
\end{align}
Further, from Eqs.~(\ref{1.32})--(\ref{1.32-1}) it
follows that
\begin{align}
    \label{a8.5}
&    \bigl[\hat{ f}_{ \mathrm{free}}  (z', \omega_1, t_1),
    \hat{ d}_A (t')\bigr] =
    -\mu_0 \omega_1 ^2\sqrt{\frac{\varepsilon_0}{\hbar \pi \mathcal{A}}}
    \,\sqrt{\varepsilon''(z',\omega_1)}
\nonumber \\&\hspace{2ex}
    \times\,
   \sum_{A'}
   \int \D t''\, \Theta(t'-t'')
   G^*(z_{A'},z', \omega_1 )
   e^{-i\omega_1 (t_1-t'')}
\nonumber \\&\hspace{15ex}
    \times\,
    \bigl[
    \hat{ d}_{A'} (t''),
    \hat{ d}_A (t')\bigr].
\end{align}
At this stage we first multiply both sides of this equation by
$i \omega_1 ^2\mu_0
\sqrt{\hbar \epsilon_0/( \pi\mathcal{A} )}\,
\sqrt{\varepsilon'' _j(\omega_1)}\,
    G^{(3j)}(0^+\!, z', \omega_1 )
%    e^{i\omega_2 (t_2-t')}
$ and perform the sum  $\sum _{j=1}^3$ and the
integrals $\int _{[j]} \D z'$, by making use of Eq.~(\ref{a4.3}).
Next we multiply the result by \mbox{$-i/(\pi \epsilon_0 \mathcal{A})
(\omega^2_2/c^2)\Theta (t_2$ $\!-$ $\!t')\mathrm{Im}\,G^{(13)}(z_{A}, 0^+\!,
\omega _2) e^{i\omega_2 (t_2 -t')}$}, take the sum with respect to $A$
and the time integral with respect to $t'$,
and recall the free-field and the source-field definitions
(\ref{1.45-1}) and (\ref{1.40}) together with
Eq.~(\ref{1.47}), leading to
\begin{align}
    \label{a8.7}
&       \bigl[
     \uh{ E}^{(3)} _{
     k{\rm free}}(0^+\!, \omega _1, t _1)
,
       \uh{ E}^{(3)\dagger} _{
     k{\rm s}}(0^+\!, \omega _2, t _2)
    \bigr]
\nonumber \\[.5ex]&\hspace{1ex}
    =
    \frac{1} {\pi ^2 \epsilon_0 ^2 \mathcal{A} ^2}
    \frac{ \omega_1 ^2 \omega_2 ^2} {c^4}
\nonumber \\[.5ex]&\hspace{2ex}
\times\,
    \sum_{AA'}
    \mathrm{Im}\,G^{(31)}( 0^+\!, z_{A'}, \omega _1)
    \mathrm{Im}\,G^{(13)} (z_{A}, 0^+\!, \omega_2 )
\nonumber \\[.5ex]&\hspace{4ex}
\times\,  \int \D t'\, \int \D t''\,
 \Theta(t_2-t') \Theta(t'-t'')
\nonumber \\[.5ex]&\hspace{6ex}
\times\,
    e^{-i\omega_1 (t_1-t'')}
    e^{i\omega_2 (t_2-t')}
    \bigl[
    \hat{ d}_{A'} (t''),
    \hat{ d}_{A} (t')\bigr].
\end{align}
Using Eqs.~(\ref{a8.8}) and (\ref{a8.7}) we then derive,
on recalling that
$\Theta(x)$ $\!+$ $\!\Theta(-x)$ $\!=$ $\!1$,
\begin{align}
    \label{a8.11}
&       \bigl[
      \uh{ E}^{(3)} _{
     k{\rm s}}(0^+\!, \omega _1, t _1)
,
       \uh{ E}^{(3)\dagger} _{
     k{\rm s}}(0^+\!, \omega _2, t _2)
    \bigr]
\nonumber \\[.5ex]&\hspace{3ex}
+
 \bigl[
      \uh{ E}^{(3)} _{
     k{\rm free}}(0^+\!, \omega _1, t _1)
,
       \uh{ E}^{(3)\dagger} _{
     k{\rm s}}(0^+\!, \omega _2, t _2)
    \bigr]
\nonumber \\[.5ex]&\hspace{3ex}
+
\bigl[
      \uh{ E}^{(3)} _{
     k{\rm s}}(0^+\!, \omega _1, t _1)
,
       \uh{ E}^{(3)\dagger} _{
     k{\rm free}}(0^+\!, \omega _2, t _2)
    \bigr]
\nonumber\\[.5ex]&\hspace{1ex}
=
-
\frac{1} {\pi ^2 \epsilon_0 ^2 \mathcal{A} ^2}
    \frac{ \omega_1 ^2 \omega_2 ^2} {c^4}
\nonumber \\[.5ex]&\hspace{2ex}
\times\,
    \sum_{AA'}
    \mathrm{Im}\,G^{(13)}(z_{A'}, 0^+\!, \omega _1)
    \mathrm{Im}\,G^{(13)} (z_{A}, 0^+\!, \omega_2 )
\nonumber \\[.5ex]&\hspace{4ex}
\times\,  \int \D t'\, \int \D t''\,
[
 \Theta(t_1-t') \Theta(t'-t'')
\Theta(t''-t_2)
\nonumber \\[.5ex]&\hspace{10ex}
+\Theta(t_2-t'')
  \Theta(t''-t')
\Theta(t'-t_1)
]
\nonumber \\[.5ex]&\hspace{6ex}
\times\,
    e^{-i\omega_1 (t_1-t')}
    e^{i\omega_2 (t_2-t'')}
    \bigl[
    \hat{ d}_{A'} (t'),
    \hat{ d}_{A} (t'')\bigr].
\end{align}

In a similar way one can
calculate the commutator
$[
      \uh{ E}^{(3)} _{k {\rm in},{\rm free}}(0^+\!, \omega, t)
,
       \uh{ E}^{(3)\dagger} _{
       k{\rm s}}(0^+\!, \omega ', t ')
    ]$, where
\begin{equation}
    \label{a8.14}
        \uh{ E}^{(3)} _{k {\rm in},{\rm free}}(z, \omega, t)
            =
            e^{-i \beta_3 z}\,
            \hat{C}
    ^{(3)}_{-} (\omega,t),
\end{equation}
with $\hat{C}^{(3)}_{-} (\omega,t)$ being given by
Eq.~(\ref{2.11}). That is, multiplying
both sides of Eq.~(\ref{a8.5}) by
$\{\mu_0  \omega_1 c
/[2 n_3(\omega_1)]\}
\sqrt{\hbar \epsilon_0/( \pi\mathcal{A} )}\,
     \sqrt{\varepsilon'' _3(\omega_1)}\,
     e^{i \beta_3(\omega_1) z'}$
and
$[i/(\pi \epsilon_0 \mathcal{A})] (\omega^2_2/c^2)
\Theta (t_2$ $\!-$ $\!t')
 \mathrm{Im}\,G^{(13)}(z_{A}, 0^+\!, \omega _2)
e^{i\omega_2 (t_2 -t')}$,
taking the integral $ \int _{[3]} \D z'$,
the time integral with respect to $t'$, and the sum with respect
to $A$, we arrive at
\begin{align}
    \label{a8.17}
&       \bigl[
      \uh{ E}^{(3)} _{
     k{\rm in},{\rm free}}(0^+\!, \omega _1, t _1)
,
       \uh{ E}^{(3)\dagger} _{
     k{\rm s}}(0^+\!, \omega _2, t _2)
    \bigr]
\nonumber \\[.5ex]&\hspace{1ex}
    =
   -\frac{i} {\pi ^2 \epsilon_0 ^2 \mathcal{A} ^2}
    \frac{ \omega_1 ^2 \omega_2 ^2} {c^4}
    \frac{\omega_1} {2c n_3}
    \int _{[3]} \D z'\,
   \varepsilon''_3(\omega_1)
\nonumber \\[.5ex]&\hspace{2ex}
\times\,
    \sum_{AA'}
     e^{i \beta_3 (\omega_1) z'}
     G^{(13)*} (z_{A'}, z, \omega_1 )
    \mathrm{Im}\,G^{(13)} (z_{A}, 0^+\!, \omega_2 )
\nonumber \\[.5ex]&\hspace{4ex}
\times\,  \int \D t'\, \int \D t''\,
 \Theta(t_2-t') \Theta(t'-t'')
\nonumber \\[.5ex]&\hspace{6ex}
\times\,
    e^{-i\omega_1 (t_1-t'')}
    e^{i\omega_2 (t_2-t')}
    \bigl[
    \hat{ d}_{A'} (t''),
    \hat{ d}_{A} (t')\bigr].
\end{align}
Now we may calculate the commutator $[
      \uh{ E}^{(3)} _{k{\rm out}, {\rm free}}(0^+\!, \omega, t)
,
       \uh{ E}^{(3)\dagger} _{
       k{\rm s}}(0^+\!, \omega ', t ')
    ]$, using the identity
\begin{align}
    \label{a8.13}
&\bigl[
      \uh{ E}^{(3)} _{k{\rm out}, {\rm free}}(0^+\!, \omega, t)
,
       \uh{ E}^{(3)\dagger} _{
       k{\rm s}}(0^+\!, \omega ', t ')
    \bigr]
\nonumber \\[.5ex]&\hspace{2ex}
=
 \bigl[
      \uh{ E}^{(3)} _{k {\rm free}}(0^+\!, \omega, t)
,
       \uh{ E}^{(3)\dagger} _{
       k{\rm s}}(0^+\!, \omega ', t ')
    \bigr]
\nonumber \\[.5ex]&\hspace{6ex}
\hspace{1ex}
-
  \bigl[
      \uh{ E}^{(3)} _{k {\rm in},{\rm free}}(0^+\!, \omega, t)
,
       \uh{ E}^{(3)\dagger} _{
       k{\rm s}}(0^+\!, \omega ', t ')
    \bigr]
.
\end{align}
Combining Eqs.~(\ref{a8.3}), (\ref{a8.7}), (\ref{a8.11}),
(\ref{a8.17}), and  (\ref{a8.13}), we derive
\begin{align}
    \label{a8.9}
&    \bigl[\hat{ b}_{k\mathrm{out}}  (t),
    \hat{b}_{k'\mathrm{out}}^\dagger (t')\bigr]
\nonumber\\[.5ex]&\hspace{2ex}
=
    \frac{1} {2\pi}
    \int_{(\Delta_k)}
    \mathrm{d}\omega\,
    \int_{(\Delta_{k'})}
    \mathrm{d}\omega '\,
    \abs{\alpha _\mathrm{out}
    }^2
    \frac{\pi\mathcal{A}}
    {\mu_0c\hbar\sqrt{\omega \omega '}}
\nonumber \\[.5ex]&\hspace{4ex}
   \times\, \bigl[
      \uh{ E}^{(3)} _{k{\rm out}, {\rm free}}(0^+\!, \omega, t)
,
       \uh{ E}^{(3)} _{k'{\rm out}, {\rm free}}(0^+\!, \omega ', t ')
   \bigr]
.
\end{align}

It can be shown \cite{khanbekyan:063812} that
the operators $\hat{ b}_{ \mathrm{out,free}}  (\omega, t)$
[and $\hat{ b}_{ \mathrm{out,free}}^\dagger  (\omega, t)$]
defined according to Eq.~(\ref{3.7}) with
$\uh{ E}^{(3)} _{\mathrm{out,free}}(z, \omega, t)$
in place of $\uh{ E}^{(3)} _{\mathrm{out}}(z, \omega, t)$
obey the Bose commutation relation
\begin{equation}
\label{3.7-1}
\bigl[\hat{ b}_{ \mathrm{out,free}}  (\omega
, t
   ),
\hat{ b}_{ \mathrm{out,free}}^\dagger (\omega'
, t'
)\bigr]
= e^{-i\omega(t-t')}\delta(\omega-\omega').
\end{equation}
Hence from Eq.~(\ref{a8.9}) it follows that
\begin{equation}
    \label{a8.9-0}
    \bigl[\hat{ b}_{k\mathrm{out}}  (t),
    \hat{b}_{k'\mathrm{out}}^\dagger (t')\bigr]
=   \delta_{kk'}
    \frac{1} {2\pi}
    \int_{(\Delta_k)}
    \mathrm{d}\omega\, e^{-i\omega(t-t')}.
\end{equation}
Extending, within the approximation scheme used,
the limits of integration to $-\infty$ and
$\infty$, we arrive at the commutation
relation (\ref{5.21}) to be proved.

%%%%%%%%%%%%%%%%%%%%%%%%%%%%%%%%%%%%%%%%%%%%%%%%%%%%%%%%%%%%%%%
\section{\label{app.7} Derivation of Eq.~(\ref{3.21}) }
%%%%%%%%%%%%%%%%%%%%%%%%%%%%%%%%%%%%%%%%%%%%%%%%%%%%%%%%%%%%%%%

To prove Eq.~(\ref{3.21}), we first show that the
integral term on the right-hand side of the Eq.~(\ref{3.20.1})
can be rewritten as follows:
\begin{align}
  \label{app.7.1}
&\hat{\Xi}_k(t) \equiv
   \int \D  t' \, \Theta ( t - t')
        e^{-i\Omega_{k}(t-t')}
\nonumber \\[.5ex]&\hspace{2ex}
\times\,
\left[
T_k(\omega)\hat{b}_{k\mathrm{in}} (\omega, t')
+\sum_\lambda A_{k\lambda}(\omega)\hat{c}_{k\lambda}(\omega, t')
\right]
\nonumber \\[.5ex]&
= \frac {1} {2\pi }
         \int_{\Delta\omega_k} \D \omega '
         \int_{t_0} ^{t + \Delta t}\! \D t''\,
          e^{i(\omega ' - \omega )(t-t'')}
  \nonumber \\[.5ex]&\hspace{2ex}
  \times\,
   \int \D  t' \, \Theta ( t - t')
        e^{-i\Omega_{k}(t-t')}
  \nonumber \\[.5ex]&\hspace{2ex}
     \times\,
\left[
T_k(\omega')\hat{b}_{k\mathrm{in}} (\omega', t')
+\sum_\lambda A_{k\lambda}(\omega')\hat{c}_{k\lambda}(\omega', t')
\right]
\end{align}
($\Delta t$ $\!\gg$ $\!\Delta _k^{-1}$).
To verify,
we first perform $t''$-integration on the right-hand
side of this equation to obtain
\begin{align}
  \label{app.7.3}
& \hat{\Xi}_k(t) =   \frac {1} {2\pi }
         \int_{\Delta\omega_k} \D \omega '
         \int \D  t' \, \Theta ( t - t')
          e^{-i\Omega_{k}(t-t')}
\nonumber \\[.5ex]&\hspace{2ex}
         \times \,
      \frac { e^{i(\omega - \omega ')\Delta t} -1 + 1-
        e^{i(\omega  - \omega ')(t_0-t)}}
    {i(\omega  - \omega ')}
\nonumber \\[.5ex]&\hspace{2ex}
     \times\,
\left[
T_k(\omega')\hat{b}_{k\mathrm{in}} (\omega', t')
+\sum_\lambda A_{k\lambda}(\omega')\hat{c}_{k\lambda}(\omega', t')
\right]
.
\end{align}
In
the coarse-grained approximation used,
i.e., $\tau$ $\!\gg$ $\!\Delta\omega_k^{-1}$
($\tau$ $\!=$ $\!\Delta t,\,t$ $\!-$ $\!t_0$),
we may let
\begin{align}
  \label{app.7.5}
&     \frac { e^{i(\omega - \omega ')\tau} -1 }
    {i(\omega  - \omega ')}
    \mapsto
    \zeta_{\pm} (\omega  - \omega ' )
\nonumber \\[1ex]&\hspace{2ex}
    =
    \left\lbrace
           \begin{array}{lll}
              i \mathrm{P}
              \displaystyle{\frac {1}{\omega  - \omega '}}
              + \pi \delta (\omega  - \omega ')
              & \ \mathrm{if}\ & \tau>0, \\[2ex]
            i \mathrm{P}
            \displaystyle{\frac {1}{\omega  - \omega '}}
              - \pi \delta (\omega  - \omega ')
              & \ \mathrm{if}\ &  \tau<0
              \end{array} \right.
\end{align}
($\mathrm{P}$, principal value). Now the $\omega '$-integration
can be easily performed to see that Eq.~(\ref{app.7.1}) is correct
within the approximation scheme used

Making
[on the right-hand side of the Eq.~(\ref{app.7.1})]
the change of variables:
$t'$
$\!\to$ $\!t$ $\!-$ $\!t''$ $\!+$ $\!t'$ and performing
$\omega ' $-integration
[$T_k$ $\!=$ $\!T_k(\omega_k)$, $A_{k\lambda}$ $\!=$
$\!A_{k\lambda}(\omega_k)$],
we find
\begin{align}
  \label{app.7.7}
&    \hat{\Xi}_k(t)=    \frac {1} {\sqrt{2\pi} }
         \int_{t_0} ^{t + \Delta t} \D t''
          \int \D  t' \, \Theta ( t'' - t')
\nonumber \\[.5ex]&\hspace{9ex}
\times\,
        e^{-i\Omega_{k}(t''-t')}
        e^{-i \omega (t-t'')}
\nonumber\\[.5ex]&\hspace{9ex}
\times
\left[
T_k\hat{b}_{k\mathrm{in}} (t')
+\sum_\lambda A_{k\lambda}\hat{c}_{k\lambda}(t')
\right]
.
\end{align}
Comparing Eq.(\ref{app.7.7}) with Eq.(\ref{2.55}), we
see that
\begin{align}
  \label{app.7.9}
   \hat{\Xi}_k(t) &=
     \left[\frac{c}{2n_1(\omega_k)l}\right]^{-\frac{1}{2}}
\nonumber\\[.5ex]&\hspace{2ex}
        \times\,
        \frac {1} {\sqrt{2\pi} }
         \int_{t_0} ^{t + \Delta t} \D t''
         e^{-i \omega (t-t'')}
          \hat{a} _k (t'') .
\end{align}
Substitution of Eq.~(\ref{app.7.9}) into Eq.~(\ref{3.20.1})
eventually yields Eq.~(\ref{3.21}).

%%%%%%%%%%%%%%%%%%%%%%%%%%%%%%%%%%%%%%%%%%%%%%%%%%%%%%%%%%%%%%%
\section
{\label{app.2.6}
Derivation of
Eq.~(\ref{6.16})
}

Using Eq.~(\ref{2.55})
together with Eqs.~(\ref{2.45})--(\ref{2.49}) and the commutation
relations (\ref{5.11}) and (\ref{5.11-1}) it is not difficult to
prove that
\begin{align}
  \label{5.51}
&  \bigl[
     \hat{a}_k  (t) ,
     \hat{b}_{k'{\rm in}}^{\dagger}  (\omega, t')
   \bigr]
         =
             \delta _{kk'}
\left[\frac{c}{2n_1(\omega_k) l}\right]^{\frac{1}{2}}
             \frac{T_k}
             {\sqrt{2 \pi}}
             \frac {i e^{-i\omega (t-t')} }
             {\omega - \Omega _k}
\,,
\\[.5ex]
  \label{5.53}
&  \bigl[
     \hat{a}_k  (t) ,
     \hat{c}_{k' \lambda}^{\dagger}  (\omega, t')
   \bigr]
        =
             \delta _{kk'}\!
\left[\frac{c}{2n_1(\omega_k) l}\right]^{\frac{1}{2}}
             \frac{A _{k\lambda}}
             {\sqrt{2 \pi}}
             \frac {i e^{-i\omega (t-t')} }
             {\omega - \Omega _k}
\,.
\end{align}
Using Eq.~(\ref{6.12}) together with Eqs.~(\ref{3.27})
and (\ref{6.8}), from Eqs.~(\ref{5.51}) and
(\ref{5.53}) we can calculate the commutator (\ref{6.16}) in a
straightforward manner.

%%%%%%%%%%%%%%%%%%%%%%%%%%%%%%%%%%%%%%%%%%%%%%%%%%%%%%%%%%%%%%%%%%%%%%%%%%%%%%

\bibliographystyle{apsrev}
\bibliography{bibl}

\begin{thebibliography}{51}
\bibitem[{tex()}]{text}
\bibinfo{note}{Note that a Hamiltonian of the form (\ref{1.3}) can be also
  obtained within the framework of a microscopic damped-polariton model
  \cite[]{suttorp:013816}}.
\end{thebibliography}

\end{document}